\documentclass{aa}
\usepackage{natbib}
\usepackage{amssymb}
\usepackage{graphicx}

\begin{document}

\title{On the nature of the extragalactic number counts in the $K$-band}
\author{
G. Barro\inst{1}\and
J. Gallego\inst{1}\and
P.G. P\'erez-Gonz\'alez\inst{1}\and
C. Eliche-Moral\inst{1}\and
M. Balcells\inst{2}\and 
V. Villar\inst{1}\and
N. Cardiel\inst{1}\and
D. Cristobal-Hornillos\inst{5}\and
A. Gil de Paz\inst{1}\and
R. G\'uzman\inst{3}\and
R. Pell\'o\inst{4}\and
M. Prieto\inst{2}\and
J. Zamorano\inst{1}}

\institute{Departamento de Astrof\'isica, Universidad Complutense de Madrid, 28040 Madrid, Spain
\and Instituto de Astrof\'isica de Canarias, V\'ia Lactea, 38200 La Laguna, Canary Islands, Spain
\and Department of Astronomy, 477 Bryant Space Center, University of Florida, Gainesville, FL
\and Laboratoire d'Astrophysique de Toulouse-Tarbes, CNRS, Universit\'e de Toulouse, 14 Avenue Edouard Belin, 31400-Toulouse, France
\and Instituto de Astrof\'isica de Andaluc\'{i}a, 18080 Granada, Spain}

\abstract{The galaxy number counts has been traditionally used to test models of galaxy evolution. However, the origin of significant differences in the shape of number counts at different wavelengths is still unclear. By relating the most remarkable features in the number counts with the underlying galaxy population it is possible to introduce further constraints on galaxy evolution.}
{We aim to investigate the causes of the different shape of the $K$-band number counts when compared to other bands, analyzing in detail the presence of a change in the slope around $K\sim17.5$.} 
{We present a near-infrared imaging survey, conducted at the 3.5m telescope of the Calar Alto Spanish-German Astronomical Center (CAHA), covering two separated fields centered on the HFDN and the Groth field, with a total combined area of $\sim0.27$deg$^{2}$ to a depth of $K\sim19$ ($3\sigma$,Vega). By combining our data with public deep $K$-band images in the CDFS (GOODS/ISAAC) and high quality imaging in multiple bands, we extract $K$-selected catalogs characterized with highly reliable photometric redshift estimates. We derive redshift binned number counts, comparing the results in our three fields to sample the effects of cosmic variance. We derive luminosity functions from the observed $K$-band in the redshift range [0.25-1.25], that are combined with data from the references in multiple bands and redshifts, to build up the $K$-band number count distribution.}
{The overall shape of the number counts can be grouped into three regimes: the classic Euclidean slope regime ($d\log N/dm\sim0.6$) at bright magnitudes; a transition regime at intermediate magnitudes, dominated by $M^{\ast}$ galaxies at the redshift that maximizes the product $\phi^{\ast}\frac{dV_{c}}{d\Omega}$; and an $\alpha$ dominated regime at faint magnitudes, where the slope asymptotically approaches -0.4($\alpha$+1) controlled by post-$M^{\ast}$ galaxies.
The slope of the $K$-band number counts presents an averaged decrement of $\sim50\%$ in the range $15.5<K<18.5$ ($d\log N/dm\sim0.6-0.30$). The rate of change in the slope is highly sensitive to cosmic variance effects. The decreasing trend is the consequence of a prominent decrease of the characteristic density $\phi^{\ast}_{K,obs}$ ($\sim60\%$ from $z=0.5$ to $z=1.5$) and an almost flat evolution of $M^{\ast}_{K,obs}$ (1$\sigma$ compatible with $M^{\ast}_{K,obs}=-22.89\pm0.25$ in the same redshift range).}{}

\authorrunning{Barro et al.}
\maketitle

\section{Introduction}\label{INTRO}
With the advent of the large photometric surveys (such as COSMOS - \citealt{2006astro.ph.12305S};
UKIDDS - \citealt{2007MNRAS.379.1599L}, etc), the coverage of the extragalactic number
counts (hereafter, NCs) on multiple bandpasses has greatly improved. Benefiting from
the new generation of wide area cameras and dedicated telescope facilities, these surveys
have provided large galaxy samples with a significant improvement in efficiency (\citealt{2008MNRAS.383.1319G}; \citealt{2003A&A...410...17M}).
Even the traditionally more problematic NIR surveys have made considerable progress in covering areas close to a square degree, up to limiting magnitudes hardly reachable a few years ago (\citealt{2006MNRAS.373L..21S}; \citealt{2008MNRAS.383.1366C}).

The NIR galaxy counts usually have been considered a particularly useful method to constrain galaxy evolution and cosmology, providing a simpler, less biased, test of galaxy evolution models. However, although the overall shape of the NCs is well defined, published counts still exhibit a considerable scatter, making it difficult to narrow down the evolution in a small magnitude interval. Additionally, the apparent simplicity of the NCs hides a mixture of evolving galaxy properties that complicates the interpretation of the observed features in terms of a single physical origin. Thus, it is not surprising that the explanation for the flatness of the $K$-band counts (relative to the optical NCs) or the presence of a sharp break in the slope around $16 < K < 18$ are still a matter of debate (\citealt{1993ApJ...415L...9G}; \citealt{2003ApJ...595...71C}; \citealt{2007AJ....134.1103Q}; \citealt{2005A&A...442..423I}; \citealt{2008A&A...482...81T}).

The change in the slope of the NCs is a direct consequence of galactic evolution. Indeed, any feature in the shape of the NCs is closely related to the luminosity distribution of galaxies at a given epoch. By disentangling the relative contribution from the luminosity functions (hereafter, LFs) at different redshift ranges, we will be able to identify the driving force behind the shape of the NCs.

Therefore, we aim to refine our understanding of the $K$-band NCs in the light of galaxy evolution by reconstructing the NCs in the $K$-band from rest-frame LFs in the multiple bandpasses probed by the observed $K$-band at different epochs. We benefit from the substantial amount of work done on deriving accurate LFs at different redshifts in the optical bands (\citealt{2003A&A...401...73W};\citealt{2003ApJ...592..819B};
\citealt{2003ApJ...586..745C}; \citealt{2005A&A...439..863I}; \citealt{2006A&A...448..101G}; \citealt{2007ApJ...656...42M}) and NIR bands (\citealt{2001ApJ...560..566K}; \citealt{2001MNRAS.326..255C}; \citealt{2003A&A...402..837P}; \citealt{2003MNRAS.342..605F}; \citealt{2007MNRAS.380..585C}; \citealt{2007arXiv0705.2438A}) to disentangle the multiple galactic populations at different redshift that assemble together to create the observed NCs, and to discern the true nature of the reported features on that distribution.

The purpose of this paper is two-fold. First we present a NIR survey conceived to serve as reference for future spectroscopic follow up. Second, we use these datasets and high quality photometric redshifts to derive NCs and LFs in the observed $K$-band, allowing us to analyze the underlying galaxy population responsible for the observed shape of the $K$-band NCs. The paper is organized as follows: In \S\ref{OBSERVATIONS} we describe the observations, data reduction and the compilation of complementary data sets. Sections \ref{CHARACTERIZATION} and \ref{THE SAMPLE} show the multi-wavelength characterization of the $K$-selected samples and the procedure to derive photometric redshifts. In \S\ref{PHOTOZ} we investigate the reliability of these photometric redshift estimates. Then, we use these $K$-selected samples in \S\ref{K_band_number_counts} and \S\ref{K_band_zphot_distrib}, presenting $K$-band NCs in three different fields and making use of the photometric redshifts to explore the redshift distribution and to probe the impact of cosmic variance in the NCs. In \S\ref{RESULTS} we discuss the connection between NCs and LFs, and present our LF estimates along with previously published results in multiple bands and redshifts. In \S\ref{NCs from LFs} we derive the expected NCs distribution from the corresponding LFs. Section \S\ref{CONCLUSIONS} summarizes the conclusions of the paper.

Throughout this paper we use Vega magnitudes unless noted otherwise and adopt the current standard cosmology $H_{0}=70$ km$^{-1}$s$^{-1}$Mpc$^{-1}$, $\Omega_{M}=0.3$ and $\Omega_{\Lambda}=0.7$.

\section{NIR observations}\label{OBSERVATIONS}
\begin{figure}[h]
\centering
\includegraphics[width=6cm]{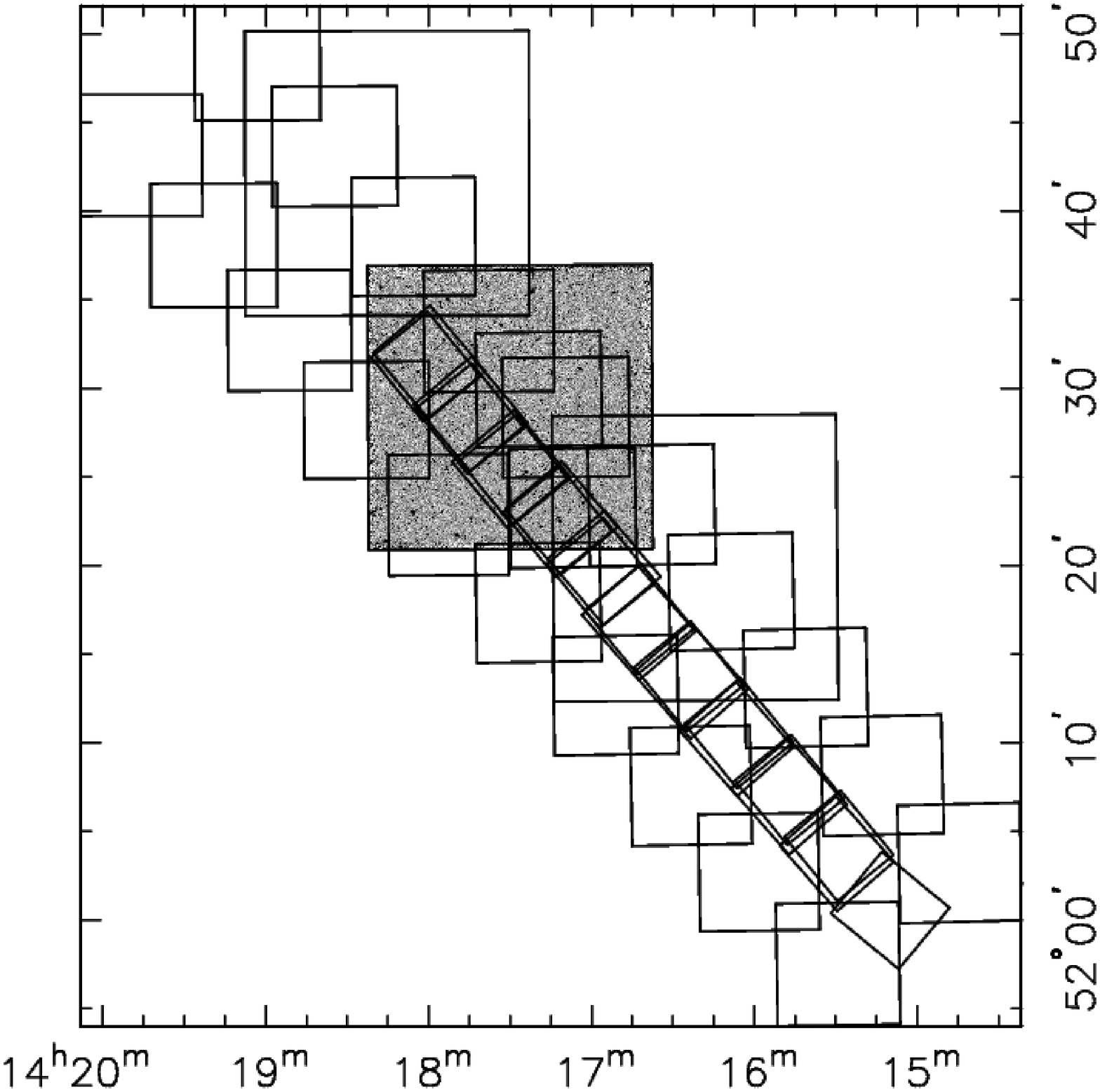}
\includegraphics[width=6cm]{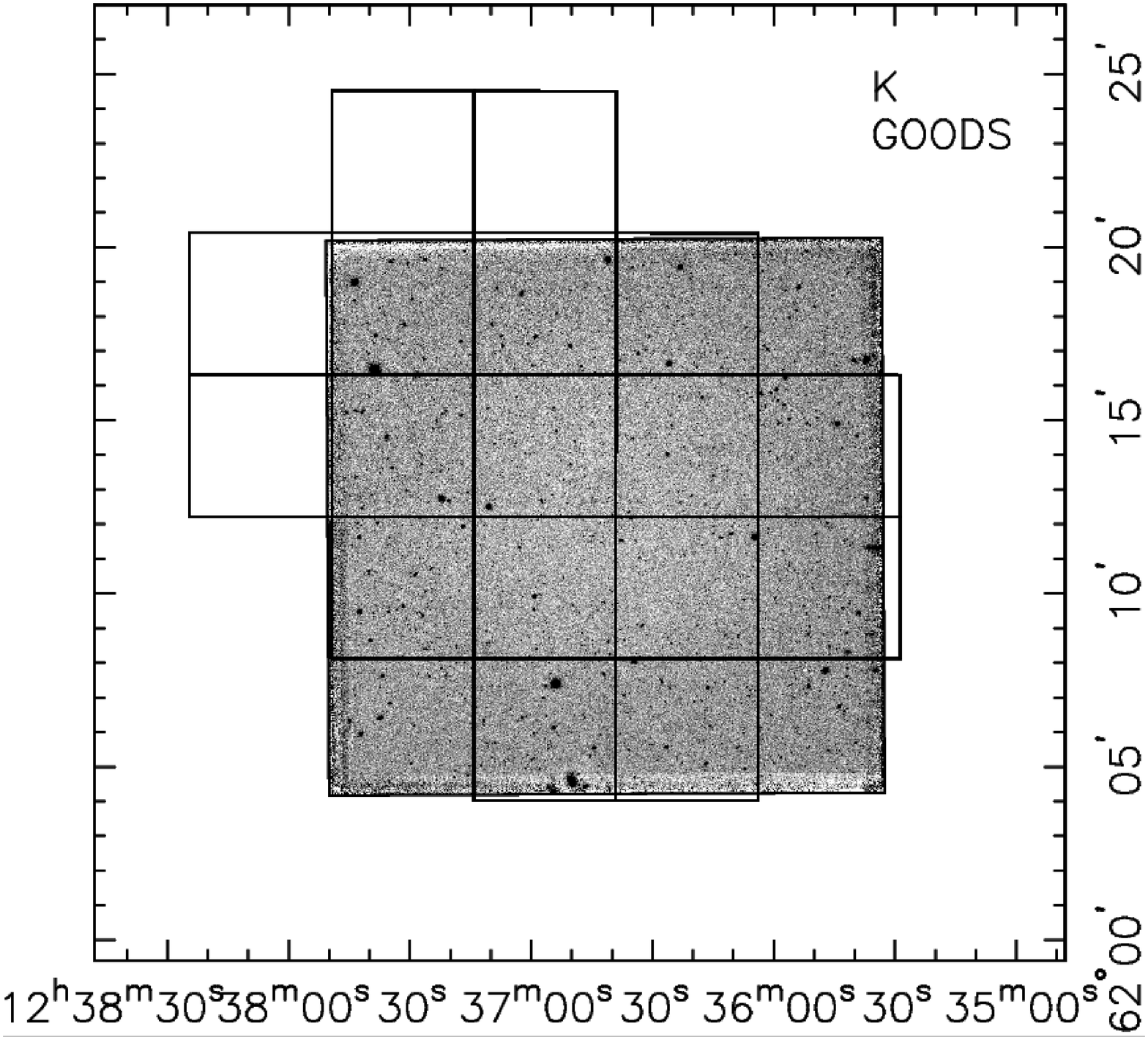}
\caption{\label{COVERAGE} Sky coverage maps of the NIR datasets in the Groth and HDFN fields. The outlines represent. \textit{Top}.- The CAHA-$\Omega'$ $K'$-band flanking fields at both sides of the WHT-INGRID $K$-band coverage of the original Groth strip \citep{2003ApJ...595...71C}, and the $\Omega2k$ $J$-band pointings of \cite{2008ApJ...677..169V}. The $J$-band central image is shown for comparison with the HDFN $K$-band image covering the same area ($\sim225$arcmin$^{2}$). \textit{Bottom}.- The CAHA-$\Omega$2k $K$-band coverage of the HDFN. The GOODS-HST footprints are also shown for comparison.}
\end{figure}

$K$-band observations were carried out on two separate fields: one
covering the flanking fields of the original Groth strip (Groth et
al. 1994) at
$\alpha(J2000.0)=14^{h}17^{m}43^{s},~\delta(J2000.0)=52^{\circ}
28'41''$ (hereafter, Groth-FF), and the second centered at the
Great Observatories Origins Deep Survey north field (HFDN;
\citealt{2004ApJ...600L..93G}),
$\alpha(J2000.0)=12^{h}36^{m}49^{s},~\delta(J2000.0)=62^{\circ}
12'58''$. Additionally, we made use of complementary infrared imaging which includes a full mapping of the Groth strip in the $J$ and $Ks$ bands (\citealt{2003ApJ...595...71C}; hereafter, CH03), and $J$-band observations in the Groth and HDFN fields (\citealt{2008ApJ...677..169V}; hereafter V08), consisting of three pointings covering at least $\sim60\%$ of the $K$ surveyed area in Groth, and a single pointing in HDFN completely overlapping with the $K$ exposure. Details of the observations are shown in Table~\ref{pointing_table}. Fig.~\ref{COVERAGE} shows the layout of the two areas covered.

The Groth-FF infrared observations in the $K'$-band were obtained using OMEGA-PRIME (hereafter, $\Omega'$), mounted on the prime focus of the 3.5m
telescope at Calar Alto Spanish-German Astronomical
Center (CAHA) during three runs in 2000 May 15-17, 2002 March 28-31 and 2002 August 19-23. The $K'$
filter was preferred to $Ks$ for a better removal of thermal
background. No significant difference in $\langle K'-Ks \rangle$ galaxy
photometry was found (see Section ~\ref{PHOTOMETRY}).

The $\Omega'$ camera was equipped a 1k$\times$1k HgCdTe Rockwell array that
provides a field of view of 7$\arcmin\times$7$\arcmin$ with $0.396''/\textrm{pixel}$ scale.
In the first runs, a small misalignment on the secondary mirror caused the PSF to be slightly
asymmetric. This problem was enhanced in the 2002 run, causing the
FWHM to be noticeably higher, and reducing the depth of the
observations. The seeing conditions were generally average, with the FWHM ranging from $1.10\arcsec$ to $1.50\arcsec$ except for a few pointings with a higher seeing ($\sim$1.9$\arcsec$). The background emission in the NIR is bright, non-uniform and highly variable. Therefore, in order to
perform an accurate sky subtraction, sky dithered exposures are
required. We used a hexagonal dithering pattern of 20$\arcsec$ side, 
repeated 12 times, shifting the center 3$\arcsec$ each time to
create the final exposure. After excluding the edges (having a lower coverage), each pointing in the Groth-FF covers $47~\textrm{arcmin}^{2}$. The
integration times were chosen to keep the count levels within the
linear regime of the detector. Typical exposure times were $31\times2$~s per
dithered position (co-adds $\times$ individual exposure time), rejecting the first frame each time to
avoid charge persistency problems. Total integration times were
around 84 minutes per pixel. Note that the integration time for the last run was increased to alleviate the alignment problem.

The whole Groth-FF mosaic comprises 22 pointings of about $7\arcmin\times7\arcmin$,
for a total of 822~arcmin$^{2}$. $\sim80\%$ of the total area is above the average exposure time per frame ($\sim80$~min), with relatively shallow limiting magnitudes of $K\sim 19.1-19.8$ ($3\sigma$). The CH03 Groth strip observations covered $\sim$180~arcmin$^{2}$ in eleven $4\arcmin\times4\arcmin$
pointings, carried out with the INGRID instrument and $K$-band filter in the William
Herschel Telescope, to a deeper limiting magnitude of
$K\sim20.5$ ($3\sigma$). The V08 central Groth strip observations
covered $702~\textrm{arcmin}^{2}$ in three 15$\arcmin\times$15$\arcmin$ pointings taken using
the OMEGA 2000 ($\Omega2k$) instrument at 3.5m CAHA to a limiting depth of $J\sim22.5$ ($3\sigma$).
The combination of the Groth NIR observations leads to a total
area of 955~arcmin$^{2}$ in the $K$-band and at least
462~arcmin$^{2}$ simultaneously observed in both the $J$ and $K$ bands.

The HDFN infrared observations in the $Ks$-band were performed with the
CAHA 3.5m telescope using $\Omega2k$, which is an improvement over
the previous $\Omega'$ instrument. The detector is a 2k$\times$2k HgCdTe
HAWAII-2 Array, with a 0.450\arcsec/pixel and a
$15.4\arcmin\times15.4\arcmin$ field of view. Observations were performed over
the course of one observing run in 2006 May 15-17. The seeing
conditions were slightly better than in Groth-FF, ranging from
$0.9\arcsec$ to $1.1\arcsec$. The typical exposure times at each dithered
position were $50\times3~s$. The total exposure time is 1.7~h. After
excluding the noisy edges, we obtained a final image of
$232~\textrm{arcmin}^{2}$ to a $K\sim19.5$ depth ($3\sigma$). The same area is covered in V08 $J$-band observations of HDFN to a limiting magnitude $J\sim21.9$ ($3\sigma$).

\subsection{Data reduction}\label{REDUCTION}

The data were reduced in a standard way using a combination of the
UCM NIR reduction software (Cardiel et al.) and the IRAF\footnotemark[1]
software package XDIMSUM. The basic methods are outlined below.

\footnotetext[1]{http://iraf.noao.edu/}
First, an average dark image with the same exposure time and number of co-adds is subtracted from the science images. Second, each science frame is flat-field corrected and the background emission is removed. Finally, all the frames are combined using an iterative method.

The skyflat image employed in the second step is built from the science frames by combining all the images with a median filter to remove sources. The background emission image is created for each science image using the median combination of the same pixel in the 3 previous images and the 3 next images.
The positions of several stars are used to determine the relative shifts between background-subtracted
images. The images are aligned to a common reference using integer pixel shifts, to preserve the Poisson nature of the noise. Since the PSF is well sampled, we do not expect an increase in the average
seeing. Object masks in the combined image are constructed using
SExtractor \citep{1996A&AS..117..393B}.
In a second step (and following iterations), the construction of the skyflat and the background estimation is repeated (and improved), this time masking out the sources detected in the previous step.

Several additional reduction procedures are carried out to improve the quality of the final images.
We create a mask of bad pixels in each image. An initial cosmetic
defect mask is created using the dark images. We then inspect each
background subtracted image individually; images with severe
artifacts, significantly higher seeing or very poor transparency are
discarded, while others with localized artifacts (e.g. satellite
trails) are masked using a custom procedure. Additional bad pixels
in each image are identified using a cosmic ray detection
procedure or are removed with a sigma-clipping algorithm during
image combination. 

\subsection{Photometric calibration}\label{PHOTOMETRY}

Photometric calibration was performed comparing aperture photometry from
Two Micron All Sky Survey (2MASS) (\citealt{2006AJ....131.1163S}) bright
stars in the final mosaic after rejecting those with poor quality flags in the 2MASS catalog.
Despite the different $K$-band filters, the integrated $K_{s}$ fluxes are essentially preserved, as
the color transformations (Eq.\ref{color_terms}) are typically smaller than a few 0.01 mag, due to the smooth
behavior of NIR SEDs.
The $\Omega'$-$K'$ and 2MASS-$K_{s}$ magnitudes are related by the following equations \citep{1992AJ....103..332W};
\begin{eqnarray}\label{color_terms}
K_{s}&=&K+0.005(J-K)\\ \nonumber
K&=&K'-(0.22\pm0.03)(H-K).
\end{eqnarray}
We find that the same equations apply for $\Omega2k$ with no significant dispersion.

The number of stars employed for each field calibration ranges
from 7 to 15, leading to a zero-point rms between 0.04 and 0.08 in
Groth-FF, and $\sim0.05$ in HDFN, which should only be
considered as a lower bound to cumulative uncertainties introduced by the color transformations. To check the quality of the Groth-FF photometric calibrations we compared them to CH03 INGRID $K$-band observations inside an $\sim$80~arcmin$^{2}$ overlapping region. Using standard stars \citep{1998AJ....116.2475P}, they found their photometric uncertainty to be less than 0.03-0.05 mag. The median offset of the comparison is less than 0.07 mag for bright objects, so we do not attempt to readjust the zero-points to their calibration.

Note that the shift in effective wavelength between the $K'$ and $K_{s}$ filters is small, and much less than the filter widths. In the discussion that follows we do not distinguish between the different $K$ filter sets.

\begin{table*}[t]
\scriptsize
\leavevmode
\begin{tabular}{cccccccc}
\hline
\hline
& R.A.&Decl&Exposure&FWHM&m(80\%eff)&$m_{lim}(5\sigma)$&Area\\
Pointing&(J2000.0)&(J2000.0)&(s)&(arcsec)&(mag)&(mag)&($arcmin^{2}$)\\
(1)&(2)&(3)&(4)&(5)&(6)&(7)&(8)\\
\hline
groth11& 14:14:43.515& +52:03:14.76& 4800& 1.15& 19.1& 19.3& 47.2\\
$groth12^{\dag}$& 14:15:28.298& +51:57:35.62& 5820& 1.53& 18.7& 19.3& 48.7\\
groth21& 14:15:11.712& +52:08:13.31& 4680& 1.20& 18.8& 19.2& 47.1\\
$groth22^{\dag}$& 14:15:56.989& +52:02:48.05& 7500& 1.50& 18.8& 19.3& 45.4\\
groth31& 14:15:39.456& +52:13:11.31& 3720& 1.50& 18.7& 19.2& 47.4\\
groth32& 14:16:21.756& +52:07:41.58& 2520& 1.30& 18.7& 19.1& 46.6\\
groth41& 14:16:06.669& +52:18:38.36& 3240& 1.16& 18.3& 18.5& 47.2\\
groth42& 14:16:49.262& +52:12:49.23& 5040& 1.31& 19.1& 19.4& 48.6\\
groth51& 14:16:35.898& +52:23:32.60& 5040& 1.00& 19.1& 19.5& 49.6\\
groth52& 14:17:17.680& +52:18:02.82& 5040& 1.14& 19.2& 19.5& 48.5\\
groth61& 14:17:07.477& +52:28:30.29& 4680& 1.30& 19.2& 19.7& 48.9\\
$groth62^{\dag}$& 14:17:50.812& +52:22:59.62& 6180& 1.50& 18.7& 19.3& 47.0\\
groth63& 14:17:04.981& +52:23:23.18& 5040& 1.14& 18.8& 19.1& 49.1\\
groth64& 14:17:17.456& +52:30:05.59& 5040& 1.07& 19.2& 19.5& 46.8\\
groth71& 14:17:35.943& +52:33:22.61& 5040& 1.42& 18.9& 19.6& 50.9\\
groth72& 14:18:21.179& +52:28:19.52& 4920& 1.80& 19.0& 19.8& 47.4\\
groth81& 14:18:03.943& +52:38:41.66& 5040& 1.30& 19.2& 19.8& 47.4\\
groth82& 14:18:49.704& +52:33:22.23& 2940& 1.90& 18.4& 19.3& 48.2\\
groth91& 14:18:33.323& +52:43:46.63& 4560& 1.50& 19.2& 19.7& 48.7\\
groth92& 14:19:17.913& +52:38:09.21& 4920& 1.60& 18.9& 19.5& 50.7\\
groth101& 14:19:02.220& +52:48:38.00& 4800& 1.90& 18.5& 19.7& 49.4\\
$groth102^{\dag}$& 14:19:45.792& +52:43:11.56& 6300& 1.50& 18.9& 19.5& 49.6\\
gooodsn& 12:36:49& +62:12:58 & 6300 & 1.1 & 19.2 & 19.5 & 232.2\\
 \hline
\end{tabular}
\caption{\label{pointing_table} Units of right ascension are hours, minutes and
seconds, and units of declination are degrees, arcminutes and
arcseconds.($\dag$) Pointing observed during secondary mirror
misalignment problem. Col. (7) is the $5\sigma$ limiting magnitude
measured inside a 1'' radius circular aperture.}
\end{table*}

\begin{figure}[b]
\centering
\includegraphics[width=9cm]{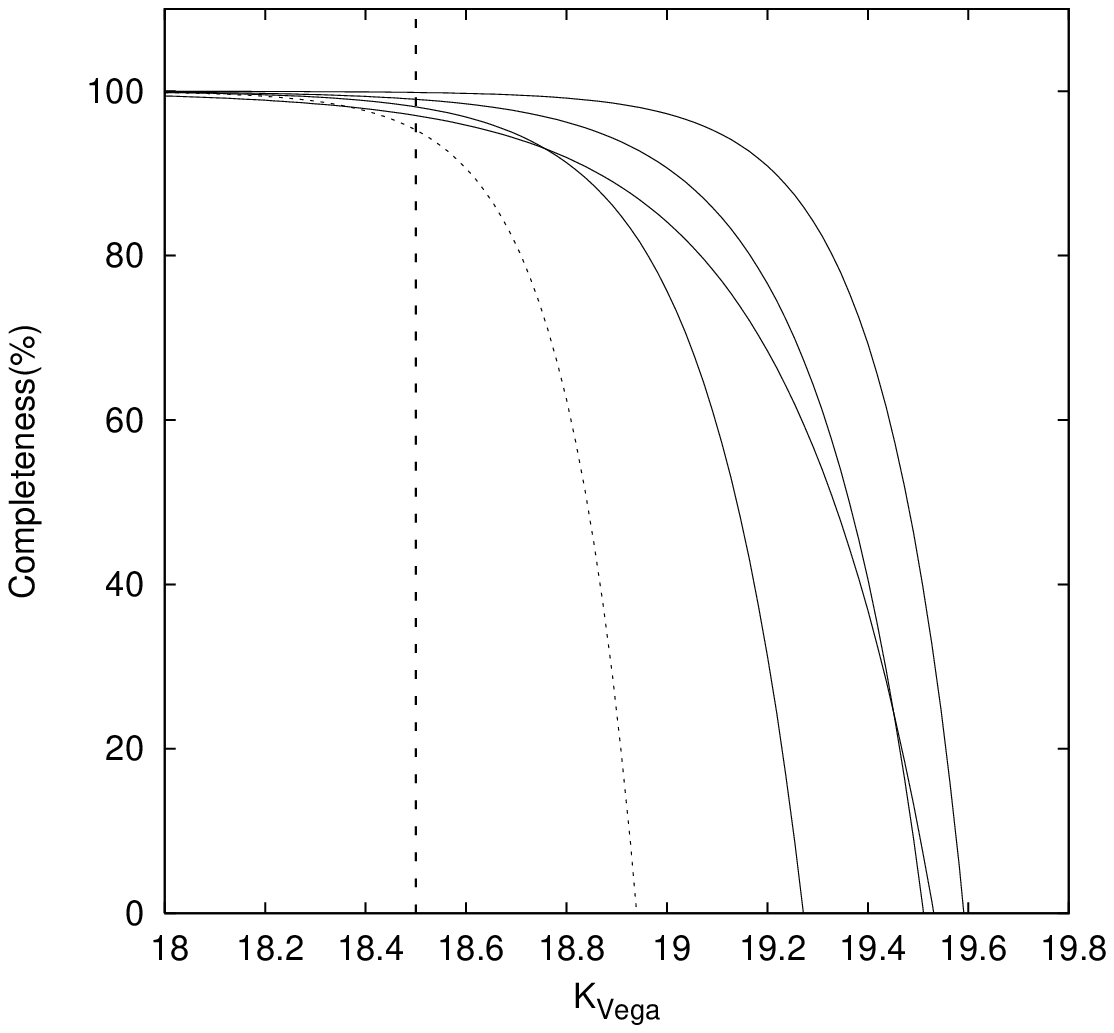}
\caption{\label{COMPLETENESS} The $K$-band best fit to the point source completeness curves. Point-like sources were inserted at random locations in the images. The completeness is defined as the fraction of recovered sources in each image. The short dashed line shows the completeness curve for the Groth32 pointing, the shallowest frame employed computing the number counts. The vertical long dashed line at $K=18.5$ depicts the magnitude threshold to which our catalog is $\sim100\%$ complete.}
\end{figure}

\subsection{Catalog completeness}\label{completeness}
We estimated the completeness of our catalogs by simulating the detection and photometry of fake sources. The fake sources were created by extracting a bright source from the image,
scaling it to the desired flux level, and injecting it at random
locations in the central well-exposed regions of the images. We
then attempt to detect these objects by running SExtractor under the
same parameters as in the original frame. Figure~\ref{COMPLETENESS} shows
the resulting completeness curves for point sources as a function of magnitude for some
of the pointings. Because the simulated sources are not required
to fall on empty regions, the confusion to real sources may slightly lower
our detection efficiency. Nonetheless, given the uncrowded nature
of our $K$-band images, we expect the confusion to affect 
our completeness values by less than $5\%$ at the $\sim80\%$ level. 

For a more realistic depth estimation, we have simulated extended sources as well as
point like sources. Extended sources are considered when the
effective radius is $20\%$ greater than the typical radius of the stars on that frame. As expected, the
efficiency for extended sources lowers the estimation from point-like sources by 0.2 to 0.3 magnitudes.
A precise determination of the completeness correction would require measuring the detection efficiency over a wider range of effective radius. Nevertheless, for the purpose of our investigation, we can establish a conservative magnitude threshold of $K$=18.5 in the HDFN and Groth-FF without affecting our conclusions. Hence, excluding the shallowest fields (groth32, groth41, groth82, groth101) from the final catalog, we obtain a galaxy sample of reasonably homogeneous depth in both fields, avoiding the use of completeness corrections.
The estimated $80\%$ completeness and limiting magnitudes for each pointing are given in Table~\ref{pointing_table}. For the Groth strip data, CH03 carried out simulations finding insignificant completeness corrections below $K=19$. We will adopt that value as the limiting magnitude.

\subsection{Complementary data}\label{MOREDATA}

\begin{table}[h]
\centering
\begin{tabular}{cccc}
\hline \hline
Band & $\lambda_{eff}(\mu m)$ & $m_{lim}$ & Source \\
(1)&(2)&(3)&(4)\\
\hline
IRAC-3.6..& 3.561  & 21.6 & \textit{Spitzer} GTO \\
IRAC-4.5..& 4.510  & 21.8 & \textit{Spitzer} GTO \\
IRAC-5.8..& 5.689  & 21.8 & \textit{Spitzer} GTO \\
IRAC-8.0..& 7.958  & 21.7 & \textit{Spitzer} GTO \\
U.........& 0.358  & 25.2 & Subaru deep $imaging^{a}$ \\
B.........& 0.442  & 25.2 & Subaru deep $imaging^{a}$ \\
V.........& 0.546  & 24.9 & Subaru deep $imaging^{a}$ \\
R.........& 0.652  & 24.4 & Subaru deep $imaging^{a}$ \\
I.........& 0.795  & 23.9 & Subaru deep $imaging^{a}$ \\
z.........& 0.909  & 23.6 & Subaru deep $imaging^{a}$ \\
b.........& 0.430  & 25.7 & $GOODS^{b}$ \\
v.........& 0.592  & 24.9 & $GOODS^{b}$ \\
i.........& 0.770  & 24.3 & $GOODS^{b}$ \\
z.........& 0.906  & 23.9 & $GOODS^{b}$ \\
$HK_{s}$ ..& 2.127 & 21.3 & QUIRC deep $imaging^{a}$ \\
b,spectra..& 0.430 & 24.4 & $TKRS^{c}$  \\
 \hline
\end{tabular}
\caption{\label{multiwav_data1} Main characteristics of the datasets in HDFN. Col.\,(1): Observing band. Col.\,(2): Effective wavelength of the filter calculated by convolving the Vega spectrum (Colina $\&$ Bohlin 1994) with the transmission curve of the filter+detector. Col.\,(3): Limiting AB magnitude. Col.\,(4): Source from where the data were obtained.
$^{a}$ Publicly available ultra-deep optical and NIR data from Capak et al.(2004).
$^{b}$ The Great Observatories Origins Deep Survey (GOODS; Giavalisco et al. 2004a).
$^{c}$ Team Keck Treasury Redshift Survey (TKRS; Wirth et al. 2004) and Cowie et al. (2004).}
\end{table}

\begin{table}[h]
\centering
\begin{tabular}{cccc}
\hline \hline
Band & $\lambda_{eff}(\mu m)$ & $m_{lim}$ & Source \\
(1)&(2)&(3)&(4)\\
\hline
Kp............&  2.114 & 20.3 & CAHA-Oprime \\
R.............&  0.652 & - & Subaru deep imaging \\
u.............&  0.381 & 27 & CFHTLS \\
g.............&  0.486 & 28.3 & CFHTLS \\
r.............&  0.626 & 27.5 & CFHTLS \\
i.............&  0.769 & 27  & CFHTLS \\
z.............&  0.887 & 26.4 & CFHTLS \\
B.............&  0.437 &  24.5 & CFHT-12k \\
R.............&  0.660 & 24.2 &  CFHT-12k \\
I.............&  0.813 & 23.5 & CFHT-12k \\
J.............&  1.255 & 22 & WHT-INGRID \\
K.............&  2.170 & 21.8 & WHT-INGRID \\
IRAC-3.6.......&  3.561 &  21.6 & \textit{Spitzer} GTO \\
IRAC-4.5.......&  4.510 &  21.8 & \textit{Spitzer} GTO \\
IRAC-5.8.......&  5.690 &  21.8 & \textit{Spitzer} GTO \\
IRAC-8.0.......&  7.957 &  21.7 & \textit{Spitzer} GTO \\
nuv...........&  0.232 & 24.5 & GALEX GTO \\
fuv...........&  0.154 & 24.5 & GALEX GTO \\
U.............&  0.361 & 24.8 & INT-WFC \\
B.............&  0.436 & 25.5 & INT-WFC \\
R,redshift....&  0.660 & 24.2 & DEEP2 \\
 \hline
\end{tabular}
\caption{\label{multiwav_data2} Main characteristics of the datasets in Groth. Col.\,(1): Observing band. Col.\,(2): Effective wavelength of the filter calculated by convolving the Vega spectrum (Colina $\&$ Bohlin 1994) with the transmission curve of the filter+detector. Col.\,(3): Limiting AB magnitude. Col.\,(4): Source from where the data were obtained.}
\end{table}

In addition to our NIR survey, high-quality imaging and photometry are publicly available for both fields. For the Groth field, we made use of some of the panchromatic data sets that have been acquired as a part of the All-wavelength Extended Groth Strip International Survey (AEGIS, see \citealt{2007ApJ...660L...1D} for a detailed data description), including ground based \textit{ugriz} deep imaging from the Canada-France Hawaii Telescope Large Survey (CFHTLS-D03-, Gwyn et al. in preparation), observed with MegaCam at the 4~m CFTH; $BRI$ wide field (0.70\degr$\times$0.47\degr) observations using the CFHT12K mosaic camera; \textit{Spitzer} mid-IR data covering the wavelength range $3.6\mu m$ to $8.0\mu m$ (GTO program) and a deep R-band image from Subaru-SuprimeCam (\citealt{2007ApJ...669..714M}).

In the HDFN, we used ultra-deep optical and NIR data spanning from the $U$ to the $HK$-band ($UBV RIzHKs$, \citealt{2004AJ....127..180C}), together with
HST-ACS \textit{bviz} imaging published by the GOODS Team \citep{2004ApJ...600L..93G}. The main characteristics of each data set are given in Table~\ref{multiwav_data1} and Table~\ref{multiwav_data2}.

At the time of writing, the CFHT12k fully reduced images were not publicly available. Instead, raw images and calibration files were downloaded from the CFHT-CADC archive and the Elixir website\footnotemark[2],
and were reduced using the IRAF MSCRED package. Precise astrometry and photometry calibrations were carried out by comparison to the CFHT12k public catalogs, and the CFHTLS and Subaru imaging in overlapping regions.

\footnotetext[2]{http://www.cfht.hawaii.edu/Instruments/Elixir/}

To complement the imaging data we have compiled a set of spectroscopic redshifts obtained by several surveys on these fields. For the HDFN we have 1699 spectroscopic redshifts from \cite{2004AJ....127.3121W}, \cite{2004AJ....127.3137C}, and \cite{2006ApJ...653.1004R}. Most of these sources are below $z\sim1$ and have a high reliability flag ($\sim80\%$). In the Groth field we have $\sim$15000 redshifts from the DEEP2 collaboration \citep{2006AAS...20919001D}. Only a small fraction of those are found within our surveyed area.

\section{Source characterization}\label{CHARACTERIZATION}
\subsection{Multi-wavelength photometry}\label{MULTIWAVELENGTH}

Multicolor photometry in all available bands was obtained employing a similar method to the one described in
P\'erez-Gonz\'alez et al. (2008, Appendix A). Briefly, the $K$-band source catalog, obtained with SExtractor, was cross-correlated in a 2$\arcsec$ radius to each one of the UV, optical and NIR catalogs. Sources detected in fewer than 3 bands were rejected. Given the relatively shallow limiting magnitude of the $K$-band images, fully covered by deep optical and IRAC images, this is a safe method to reject spurious sources. Then, aperture matched photometry was performed using elliptical Kron-like apertures. The aperture size and orientation is determined by the $K$-band image and translated to all the other optical and NIR images except for the IRAC bands.

For our typical seeing values the aperture is large enough to enclose the PSF in all these bands. For the IRAC bands, where the resolution is slightly worse ($\sim$2\arcsec; \citealt{2004ApJS..154...39F}), the flux was measured in small circular apertures (typically 3\arcsec) and corrected using stellar PSF growth curves (similarly to \citealt{2004ApJS..154...44H}; \citealt{2008arXiv0803.0748B}). In addition, we have also applied a deblending algorithm for the sources with multiple $K$-band counterparts associated with a single IRAC source and separated by more than $1\arcsec$ (the approximate limit of the astrometric resolution). The $K$-band positions of the sources were used to re-align the position of the aperture in the IRAC band. Then, the IRAC PSF is convolved with the $K$-band PSF and the flux is measured again inside a $0.9\arcsec$ aperture, applying the corresponding aperture correction. For source separations larger than $1\arcsec$, the flux contamination with this method is lower than $10\%$ \citep{2008ApJ...675..234P}.

For GALEX and HST data, we took the SExtractor MAG\_BEST magnitude of the closest source. Consequently, we do not use these data in the photometric redshift determination. For the rest of the bands, the procedure allows us to obtain accurate colors, since we measure a similar fraction of the flux from an object in each filter.

Uncertainties in the measured flux were derived taking into account the background noise, photon statistics, readout noise and scatter in the WCS. The standard method of determining the background noise from pixel-to-pixel variations of adjacent sky pixels often underestimates the real value due to correlated signal effects introduced during the data reduction (\citealt{2003AJ....125.1107L};\citealt{2006ApJS..162....1G}). To obtain a more accurate determination of the background noise we followed the method described in \cite{2008ApJ...675..234P}, which is similar to that of \cite{2003AJ....125.1107L}. For each source, the sky flux was measured in randomly distributed apertures with the same size as the photometric aperture. Then, we estimated the background noise from the width of the flux histogram, approximated by a Gaussian distribution. The comparison to the sky background measured with SExtractor shows that the latter tends to underestimate (10$\%$-15$\%$) the noise even in the $K$-band images, where integer pixel shifts were used during the frame combination.

\subsection{Photometric redshift}

We calculate photometric redshifts using the methods described in detail in Per\'{e}z Gonz\'{a}lez  et al. (2005,2008). Briefly, a reference set of galaxy templates is fitted to the observed spectral energy distributions (SEDs), taking into account the flux uncertainties. The galaxy templates are composed of a sample of galaxy spectra from HDFN and CDFS with highly reliable redshift determinations, including some $z>1.5$ galaxies, and well-covered SEDs (with more than 10 different photometric data points). The set is fitted with models of single, and composed stellar populations (1-POP and 2-POP) and models of dust emission. The comparison between data and models is done using a maximum likehood estimator that takes into account the uncertainties in each data point. The resulting 1-POP templates are characterized by four parameters: The star formation time scale $\tau$, age \textit{t} (assuming exponentially declining star formation history), metallicity $Z$, and extinction A(V). For the 2-POP we have twice the same family of parameters to characterize both the young (instantaneous burst) and the old populations. The final set comprises 3624 models (1666 1-POP+dust, 1958 2-POP+dust). The photometric redshift estimation and uncertainty is derived from the minimization of the $\chi^{2}$  probability distribution obtained from the fitting of the observed data to a grid of redshifted models (using $\delta z=0.01$). In addition, we include AGN templates from \cite{2007ApJ...663...81P} to provide a better fit of the very few AGN dominated SEDs.

Furthermore, it is possible to use the best fitting SED to estimate rest-frame fluxes, luminosities and also observed fluxes in bands that might be missing due to shallow imaging or differences in the covered area. The use of these synthetic magnitudes allows a consistent calculation of colors for all sources in the sample.

\subsection{Star-galaxy separation}\label{STAR_GALAXY}

\begin{figure}[h]
\centering
\includegraphics[angle=0, width=8cm]{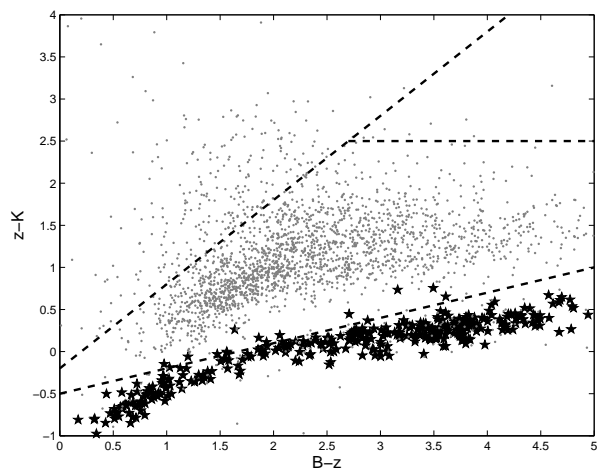}
\caption{\label{STELLARITY} Efficiency of the BzK color-color diagram to identify stars, compared to the combination of the other 9 criteria applied to isolate stars. The dashed lines depict the regions defined by the BzK criteria. The bottom line isolates the stellar locus, while the other two select sources at z$>$1.4. Black stars show the sources that satisfy three or more stellarity criteria.}
\end{figure}

Stars are separated from galaxies using a combination of morphological and color criteria. Following P\'erez-Gonz\'alez et al. (2008), all secured objects (with detection in more than 3 photometric bands) are classified as stars if they satisfy at least 3 of 9 morphological and color criteria. The first one is the all-band weighted SExtractor STELLARITY parameter. When available, the STELLARITY estimations from HST images are given a higher weight in the comparison. The other criteria are based on nIR and IRAC color-color criteria and comparisons between magnitudes derived from isophotal and circular apertures extracted from \cite{2004ApJS..154...48E} and \cite{2005AJ....129.1183R} (see P\'erez-Gonz\'alez et al. 2005, 2008 for a detailed description). We also consider the BzK criteria by \cite{2004ApJ...617..746D}, to isolate the stellar locus: $(z-K)_{AB} < 0.3\times(B-z)_{AB}-0.5$. In Fig.~\ref{STELLARITY}  we compare the efficiency of the latter to the other color-morphology criteria. Sources classified as galaxies are required to satisfy fewer than three different criteria. We find that almost all the stars satisfy the BzK criteria with only a few objects below our star threshold, probably being galaxies scattered into the stellar boundary due to photometric errors as already reported by other authors \citep{2006ApJ...638...72K}.

We also classify as stars a very small fraction of objects whose best fit to a spectral template is the Vega SED, and that does not have a reliable spectroscopic redshift. These objects typically satisfy at least 2 of the other criteria, and after visual inspection, it is most likely that they are stars.

\section{Composite sample}\label{THE SAMPLE}

Our final K-selected sample is comprised of 1313 sources in the 223~arcmin$^{2}$ of HFDN, 4660 sources in the 648~arcmin$^{2}$ of the Groth-FF, and 1957 sources in the 161~arcmin$^{2}$ of Groth strip. Out of these, less than $15\%$ are identified as stars in each field. We merged both Groth samples into a single catalog with 5948 sources. To avoid repeated sources, the two samples were cross-correlated in a 1$''$ radius removing duplicate sources with lower S/N.

Additionally, to complement the low depth samples, we have created a K-band selected sample in the Chandra Deep Field South (CDFS, $\alpha(J2000.0)=03^{h}32^{m}28^{s},\delta(J2000.0)=-27^{\circ} 48'30''$).
The reference image was drawn from the latest data release of the GOODS/ISAAC observations (Retzlaff et al. in preparation). This final release adds two new frames in the K-band, increasing the total surveyed area to 172~arcmin$^{2}$. The combined mosaic covers the region with a variety of exposure times, with an average depth of $K\sim22.7$ and a maximum of $K\sim24.2$. Consequently, we can safely consider the whole sample to be complete below $K=20$ (see also the area-depth estimations from \citealt{2006A&A...449..951G} for the v1.5 data). The CDFS has also been the target of extensive multi-wavelength observations. We make use of all the public and proprietary data compiled in \cite{2005ApJ...630...82P} to create the multicolor sample, following the procedures described in section \S\ref{CHARACTERIZATION}. After excluding a small portion of the mosaic with very low S/N regions, the sample contains 6810 galaxies in an area of 168~arcmin$^{2}$.

\section{Photometric redshift distribution}\label{PHOTOZ}

\begin{figure}[t]
\includegraphics[angle=0, width=8cm]{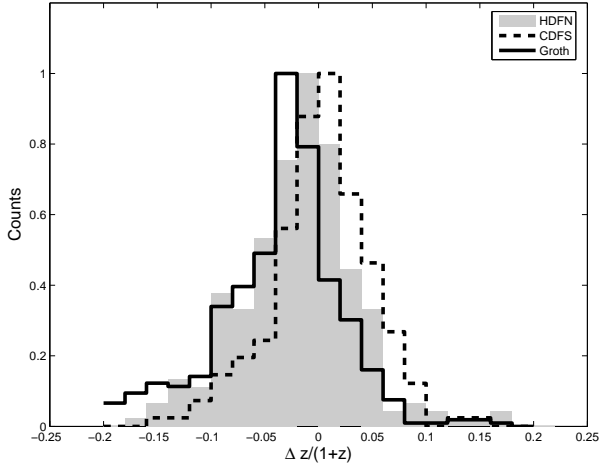}
\caption{\label{zspec_vs_zphot} Comparison of the scattering in the photometric redshift estimates ($z_{\mathrm{spec}}-z_{\mathrm{phot}})/(1+z_{\mathrm{phot}}$) for the three fields.}
\end{figure}

We have checked the quality of our photometric redshifts with all the available spectroscopic data compiled for the three fields (Fig.~\ref{zspec_vs_zphot}). For HDFN there are 287 sources ($36\%$ of the sample) with highly reliable spectroscopy. The average redshift difference ($\delta z=z_{\mathrm{spec}}-z_{\mathrm{photo}}$) is $0.001$, $70\%$ of the sample present values of $\sigma_{z}/(1+z)<0.05$ (where $\sigma_{z}$ is the absolute value of $\delta z$), and $91\%$ have $\sigma_{z}/(1+z)<0.1$. In the CDFS 232 sources ($53\%$ of the sample) have spectroscopic redshift with a high quality flag. The mean value of $\delta z$ is $0.014$. $72\%$ of the sample present values of $\sigma_{z}/(1+z)<0.05$ and $91\%$ have $\sigma_{z}/(1+z)<0.1$. In the Groth-FF we find 846 sources with high quality flag ($35\%$). The average $\delta z=0.036$. $59\%$ of the sample present values of $\sigma_{z}/(1+z)<0.05$ and $85\%$ have $\sigma_{z}/(1+z)<0.1$. Finally, there are 334 spectroscopically measured sources in the Groth strip with an  average $\delta z=0.010$. The $68\%$ and $89\%$ have values of $\sigma_{z}/(1+z)$ below $0.05$ and $0.10$ respectively.

\begin{figure}
\centering
\resizebox{0.90\hsize}{!}{\includegraphics[angle=0]{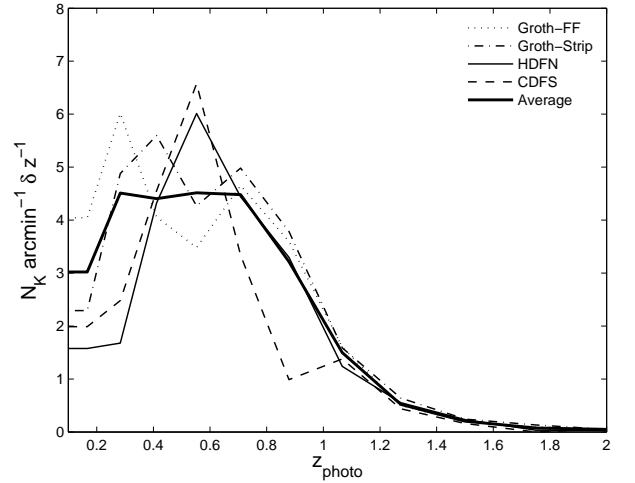}}
\caption{\label{zphot_distrib}Photometric redshift distributions of the $K$-band selected galaxies in the HDFN, CDFS and Groth fields.}
\end{figure}

Figure~\ref{zphot_distrib} shows the photometric redshift distribution for the samples in Groth, HDFN and CDFS to a limiting magnitude of $K=18.5$. The distributions have been derived taking into account the typical photometric redshift error. Although the shape of the distributions are typical for a flux limited sample, field-to-field differences reveal the effects of large scale structure. The peak of the redshift distribution depends on the luminosity function of the galaxies at different redshifts. At $K<18.5$ the majority of the sources are found at $z < 0.7$ (see next section), leading to an exponentially decreasing tail at higher redshifts. The prominent density peak in HDFN at $z\sim0.5$ is in good agreement with the redshift distribution found by the Team Keck spectroscopic survey \citep{2004AJ....127.3121W}. The less pronounced feature in CDFS at approximately the same redshift also coincides with a spectroscopically confirmed overdensity at $z=0.7$ \citep{2006A&A...454..423V}, if we take into account the broadening caused by the photometric redshift uncertainties and the decreasing selection function at high redshift. Finally, there appears to be a slight underdensity at $z\sim0.4$ and a small peak at $z\sim0.2$ in the Groth field. This alternation of peaks and voids between fields highlights the impact of field-to-field variance in the small area surveys, especially at $z\leq1$.

\section{K-band number counts}\label{K_band_number_counts}

\begin{figure}
\centering
\resizebox{0.99\hsize}{!}{\includegraphics{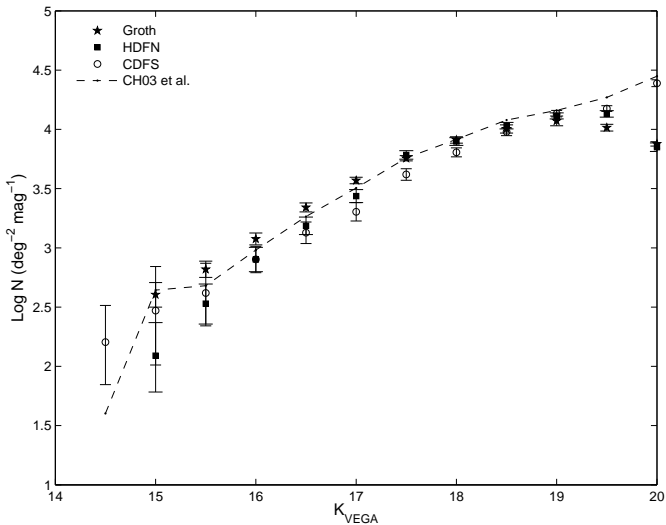}}
\caption{\label{counts3fields} The raw number counts in our three fields (HDFN, CDFS, Groth) and counts from CH03 obtained in an small portion of our total area in the Groth field. Error bars are derived from a combination of Poisson errors and bootstrapping. No completeness corrections have been made.}
\end{figure}

Figure~\ref{counts3fields} shows the $K$-band NCs for the three fields, as well as the  NCs from CH03, for comparison. The NCs in 0.5 magnitude bins, up to $K$=18.5 for Groth and HDFN, and up to $K$=20.0 for CDFS are summarized in Table \ref{K_NUMBER_COUNTS}.  Within these limits, detection efficiencies for point sources are well above 90\% (see Fig.~\ref{COMPLETENESS}, and completeness estimates in \S\ref{completeness} and \S\ref{THE SAMPLE}).  Hence, we do not apply completeness corrections to the NCs.  Our counts do not need corrections for spurious sources either; spurious sources typically appear at brightness levels where the efficiency drops
significantly below 100\% (CH03;\citealt{2006ApJ...639..644E}), and the fraction of spurious sources in our catalogs must be negligible given that each source is required to be detected in at least three bands (\S\ref{MULTIWAVELENGTH}). NCs are tabulated in Table~\ref{K_NUMBER_COUNTS}. The error calculation assumes Poisson statistics for low numbers \citep{1986ApJ...303..336G} added in quadrature to the standard deviation derived from bootstrapping the source magnitudes convolved with a Gaussian error kernel. 

\begin{table}[t]
\centering
\small
\begin{tabular}{cccc}
\hline
& Groth & HDFN & CDFS \\
\hline
$K$ Bin Center & log(N) & log(N) & log(N) \\
\hline
16.00 &$3.07^{0.05}_{0.05}$ & $2.90^{0.10}_{0.10}$ & $2.90^{0.11}_{0.11}$ \\
16.50 &$3.34^{0.04}_{0.04}$ & $3.18^{0.07}_{0.07}$ & $3.13^{0.10}_{0.09}$ \\
17.00 &$3.56^{0.03}_{0.03}$ & $3.44^{0.06}_{0.05}$ & $3.30^{0.08}_{0.08}$ \\
17.50 &$3.75^{0.02}_{0.02}$ & $3.78^{0.04}_{0.04}$ & $3.62^{0.05}_{0.05}$ \\
18.00 &$3.91^{0.03}_{0.03}$ & $3.89^{0.03}_{0.03}$ & $3.81^{0.04}_{0.04}$ \\
18.50 &$4.00^{0.03}_{0.03}$ & $4.03^{0.03}_{0.03}$ & $3.97^{0.03}_{0.03}$ \\
19.00 &       -             &          -           & $4.13^{0.03}_{0.03}$ \\
19.50 &       -             &          -           & $4.18^{0.02}_{0.02}$ \\
20.00 &       -             &          -           & $4.38^{0.03}_{0.03}$ \\
\hline
\end{tabular}
\caption{\label{K_NUMBER_COUNTS} Differential number counts in 0.5 magnitude bins for the $K$-selected samples in HDFN, CDFS and the Groth field, uncorrected for completeness. Only the NCs up to the completeness limit of each field are shown.}
\end{table}

\begin{figure*}
\resizebox{0.48\hsize}{!}{\includegraphics{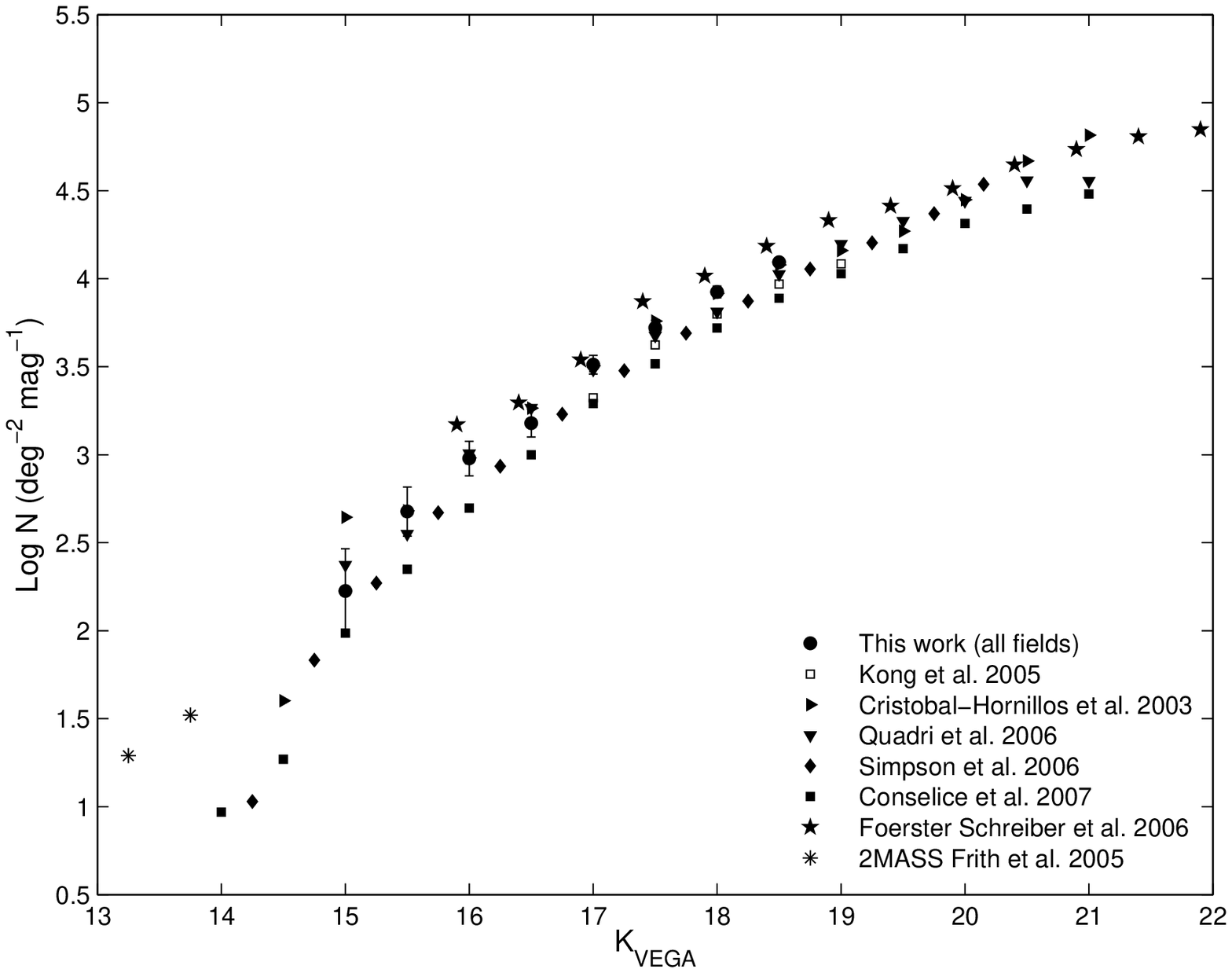}}
\resizebox{0.48\hsize}{!}{\includegraphics{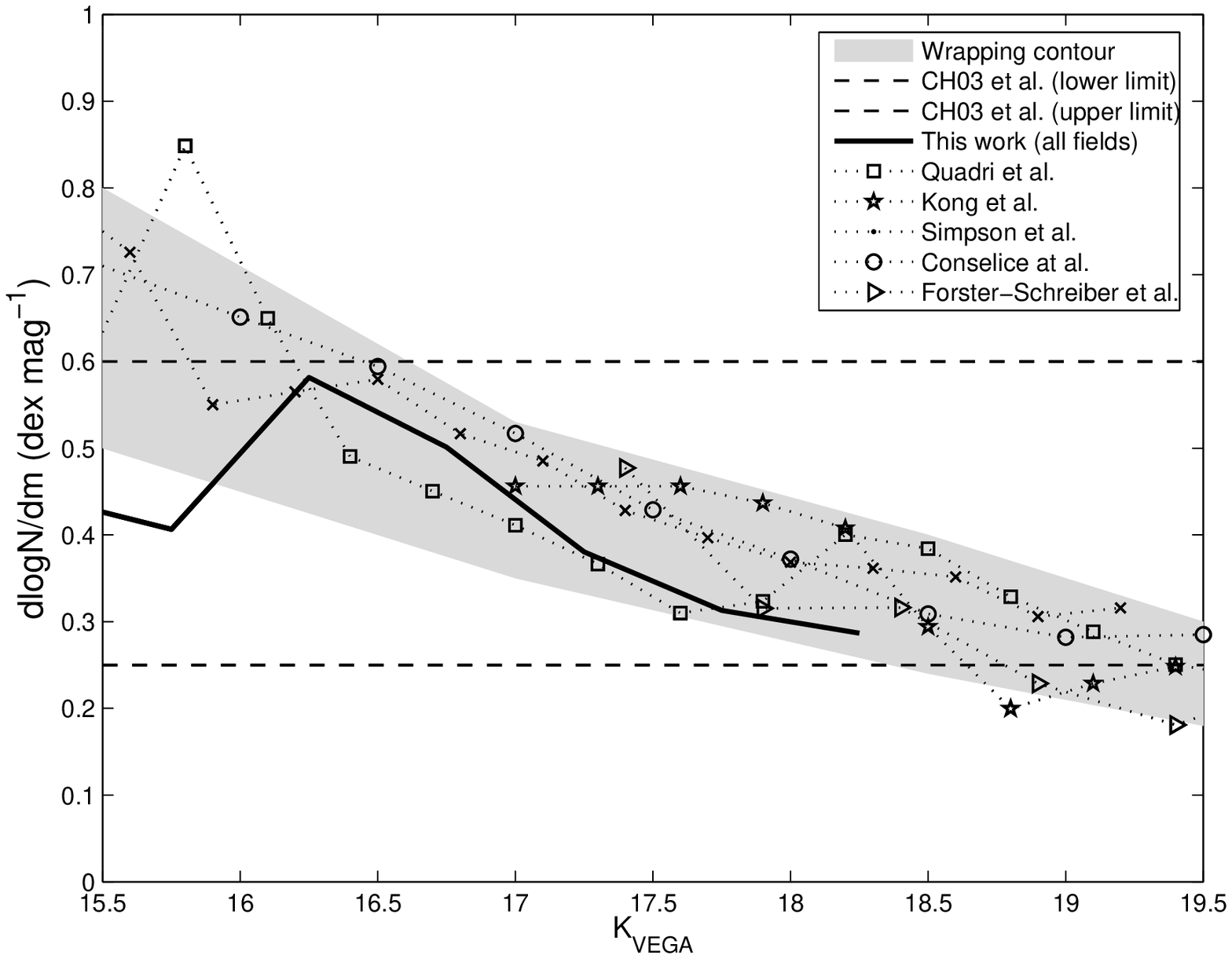}}
\caption{\label{counts_slope_fields} \textit{Left}: Averaged
$K$-band number counts from Groth, HDFN and CDFS compared with a
compilation of results taken from various sources. \textit{Right}:
Slope for the differential number counts in our three fields,
along with data from the literature. The horizontal dashed lines indicates the value of the slope at both sides of $K=17.5$ reported in CH03.}
\end{figure*}

In the Groth field, our counts are in good agreement with those of CH03. Differences are 1$\sigma$ compatible in the range 16.0$<$$K$$<$ 18.5. The overall agreement provides an external check on the quality of both sets of $K$-band number counts in the Groth field. As we move to brighter magnitudes, our counts are systematically above those of CH03 and HDFN. Such an offset is likely due to the smaller area mapped in these two surveys ($\sim$5 times smaller in the case of CH03). Nevertheless, NCs below $K<16$ probe very low redshift populations (see next section) that would require larger areas to be studied accurately. 

The NCs in the Groth and HDFN fields at $K>18.5$ are shown in Fig.~\ref{counts3fields} for comparison to the deeper CDFS counts and the completeness corrected counts of CH03. The fact that the shallower NCs do not fall abruptly until $K\sim$19 supports our efficiency estimates for the Groth and HDFN images. In addition, it is noteworthy that the scatter in the brighter counts ($K<17$) is on average greater than the typical uncertainties for the NCs in any of the fields. The most plausible cause of this discrepancy is cosmic variance. As we showed in the previous section (Fig.~\ref{zphot_distrib}), there might be significant overdensities at some redshifts that lead to these differences. Additionally, the effect is accentuated by the fact that bright counts come mainly from low redshift sources, which are poorly sampled in small area surveys, like HDFN and CDFS. We will further discuss field-to-field variance effects in the next section.

The left panel of Fig.~\ref{counts_slope_fields} shows the averaged NCs of the 3 fields along with counts from the literature. We have compiled measurements from large area surveys to mitigate the effects of cosmic variance on the shape of the NCs. That is the case of the 1.47~deg$^{2}$ of the DEEP2/Palomar survey \citep{2008MNRAS.383.1366C}, the 0.70~deg$^{2}$ of the UKIDDS-UDS \citep{2006MNRAS.373L..21S}, the 0.17~deg$^{2}$ of the Daddi-F \citep{2006ApJ...638...72K} and the combined $\sim$0.11~deg$^{2}$ of the MUSYC survey \citep{2007AJ....134.1103Q}. Additionally, we compared our results with the deeper NCs of \cite{2006AJ....131.1891F} in the MS1054-03 galaxy cluster to illustrate the asymptotic behavior of faint counts and also the impact on the NCs of a very prominent overdensity such as a cluster at z=0.83. It can be seen in Fig.~\ref{counts_slope_fields} that the density peak dominates the bright counts. This is comparable, to a lesser extent, to the effect at $K<17$ of the low redshift peak in CDFS.

\begin{figure}
\centering
\resizebox{0.90\hsize}{!}{\includegraphics{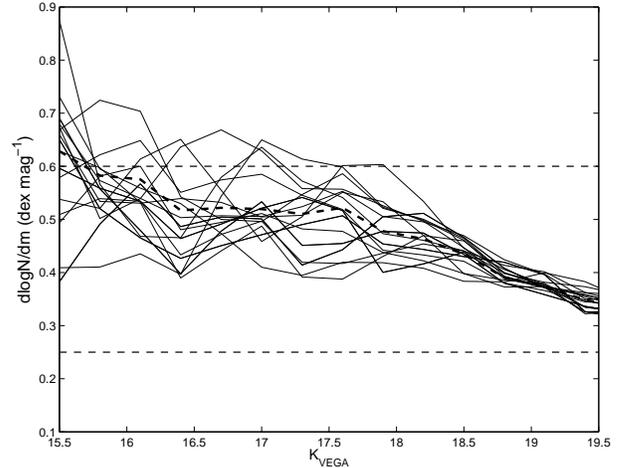}}
\caption{\label{counts_slope_millenium}Slope for the differential number counts in the 100 samples of the Millennium simulation drawn from the mock catalogs of \cite{2007MNRAS.376....2K}. The thick dashed line shows the averaged NCs combining all the samples.}
\end{figure}

The logarithmic derivative of the differential NCs (i.e., the slope of the NCs) is shown in the right panel of Fig.~\ref{counts_slope_fields}. The horizontal lines at 0.25 and 0.6 indicate the value of the slope on both sides of $K=17.5$ reported in CH03. The thick black line shows the evolution of the slope for the combined NCs of our three fields. We confirm that the logarithmic slope decreases rather sharply from 0.6 at $K=16.25$ to $\sim$0.3 at $K=$18.0, a similar result to CH03. However, it appears that the evolution of the slope is best described by a continous decreasing trend, rather than by fixed values on both sides of $K\sim17.5$; i.e., the slope of $K$-band NCs decreases monotonically over the 16.0$<K<$19.0 range. The logarithmic slopes derived from counts by other authors are bound by the same upper and lower limits as in our data, except at $K < 16.0$. However, they show a more gentle variation with magnitude, as well as a greater dispersion.  Such variations among different surveys may be driven by two processes. First, completeness and reliability of the photometry are treated with varying degrees of rigor by different authors, which contributes to the scatter in the NCs (see CH03). Second is cosmic variance.  Substructure in the redshift distribution can cause significant fluctuations in the characteristic density, and hence slope variations, that might lead either to a sharp break or to a smoother evolution in the slope of NCs. These density peaks are clearly recognizable in the right panel of Fig.~\ref{counts_slope_fields} at $K=16.5$ and $K\sim18$ when comparing the NCs of HDFN and CDFS. To strengthen this idea we have created 100 Millennium simulation mock catalogs from \cite{2007MNRAS.376....2K} selected over a 0.25~deg$^{2}$ area, similar to that of our Groth field, to sample the extent of field-to-field variance effects. Although the $K$-band NCs from \cite{2007MNRAS.376....2K} do not  accurately reproduce the observed distribution at $K>17$ (mainly because the $z>1$ galaxy population is overestimated compared to the observations), their results on clustering analysis are consistent with observations. Thus, the simulation is suitable to mimic the effects of cosmic variance. As can be seen in Fig.~\ref{counts_slope_millenium}, the broadening effect in the slope of the bright counts is similar to ours, whereas the confidence interval for the faint counts is significantly narrower, probably because at that point our NCs are dominated by photometric errors and depth effects, and not by large scale structure (LSS).

\section{Redshift distribution of K-band number counts}\label{K_band_zphot_distrib}

The change in the slope of NCs is an indirect effect of galactic evolution. Indeed, any feature in the shape of NCs is closely related to the luminosity distribution of galaxies at a given epoch. By disentangling the relative contribution to the NCs of the LFs at different redshift ranges, we will be able to identify the main effect responsible for shape of the NCs.

\begin{figure}
\centering
\resizebox{0.9\hsize}{!}{\includegraphics{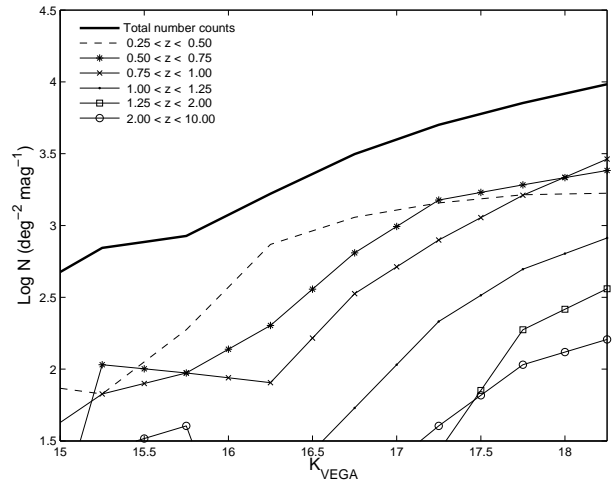}}
\caption{\label{counts_inbins} NCs in the $K$-band for the CDFS field alone (thick black line) and NCs in several photo-redshift ranges. At bright magnitudes ($K$$\sim$17) the contribution to the NCs is limited to only a few low redshift bins, whereas at $K$$\sim$18 the total number counts becomes a mixture of several low and high redshift bins ($z>1.5$)}
\end{figure}

Fig.~\ref{counts_inbins} depicts the $K$-band NCs in the CDFS, the deepest sample. We have used the photometric redshift estimation to split the total number counts into redshifts bins. The binned NCs tend to resemble the shape of a (Schechter) LF. The evolution of the comoving distance with redshift causes an effective shift in the successive redshift bins towards fainter magnitudes, and the posterior accumulation of the higher redshift NCs. In addition, despite the likely greater effects of cosmic variance on a single field, it seems clear that for $K>18$ the total number counts tend to become a mixture of very different redshifts ranges, with a growing contribution of high redshift galaxies ($z>$1.5) absent at brighter magnitudes, where 2-3 low redshift bins account for $90\%$ of the total NCs \citep{2008MNRAS.383.1366C}. Furthermore, it can be seen that for $K<16$ the counts sample the bright part of the LFs at $z<$1, causing values to be highly sensitive to both volume effects and LSS.

\begin{figure*}
\centering
\resizebox{0.38\hsize}{!}{\includegraphics{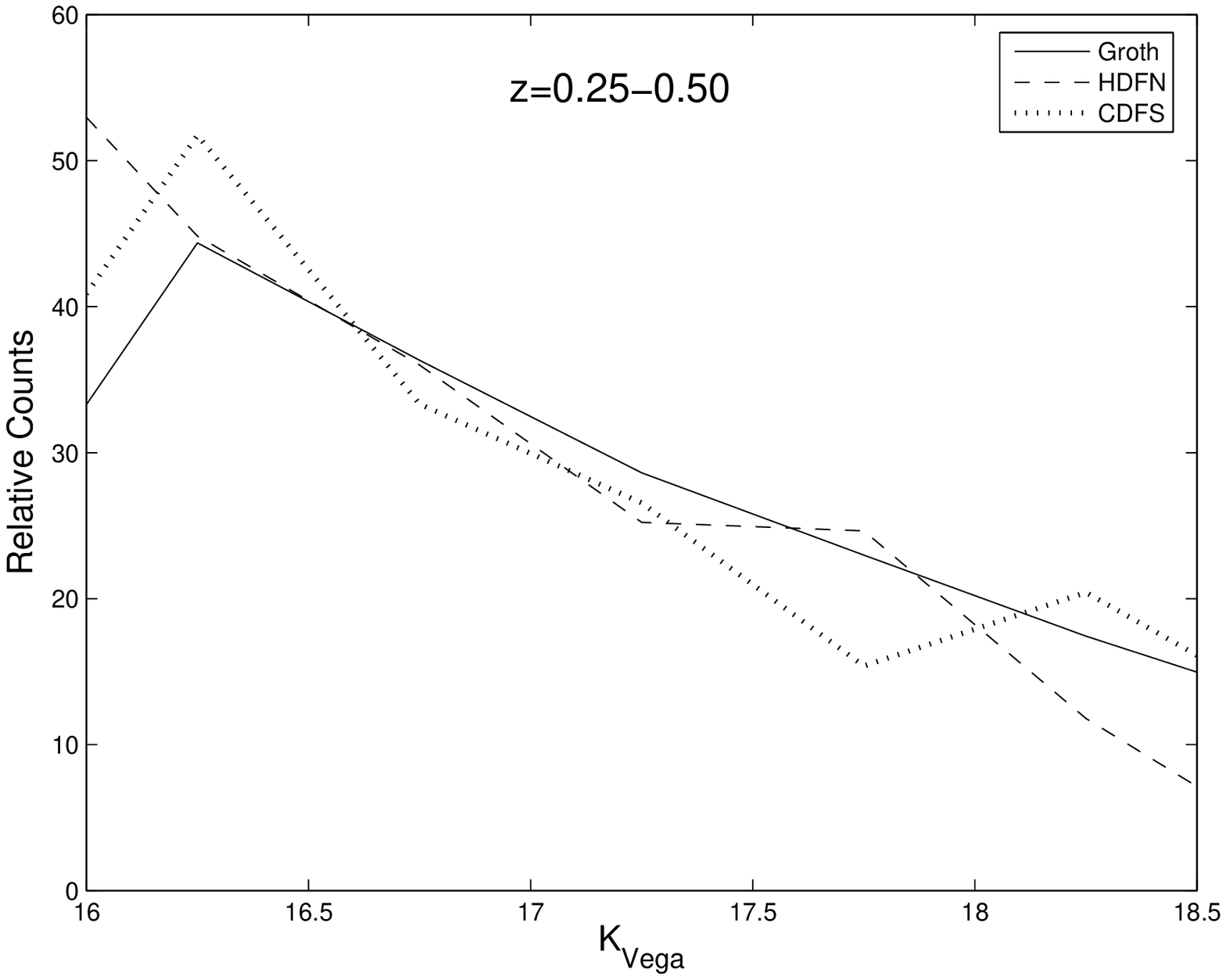}}
\resizebox{0.38\hsize}{!}{\includegraphics{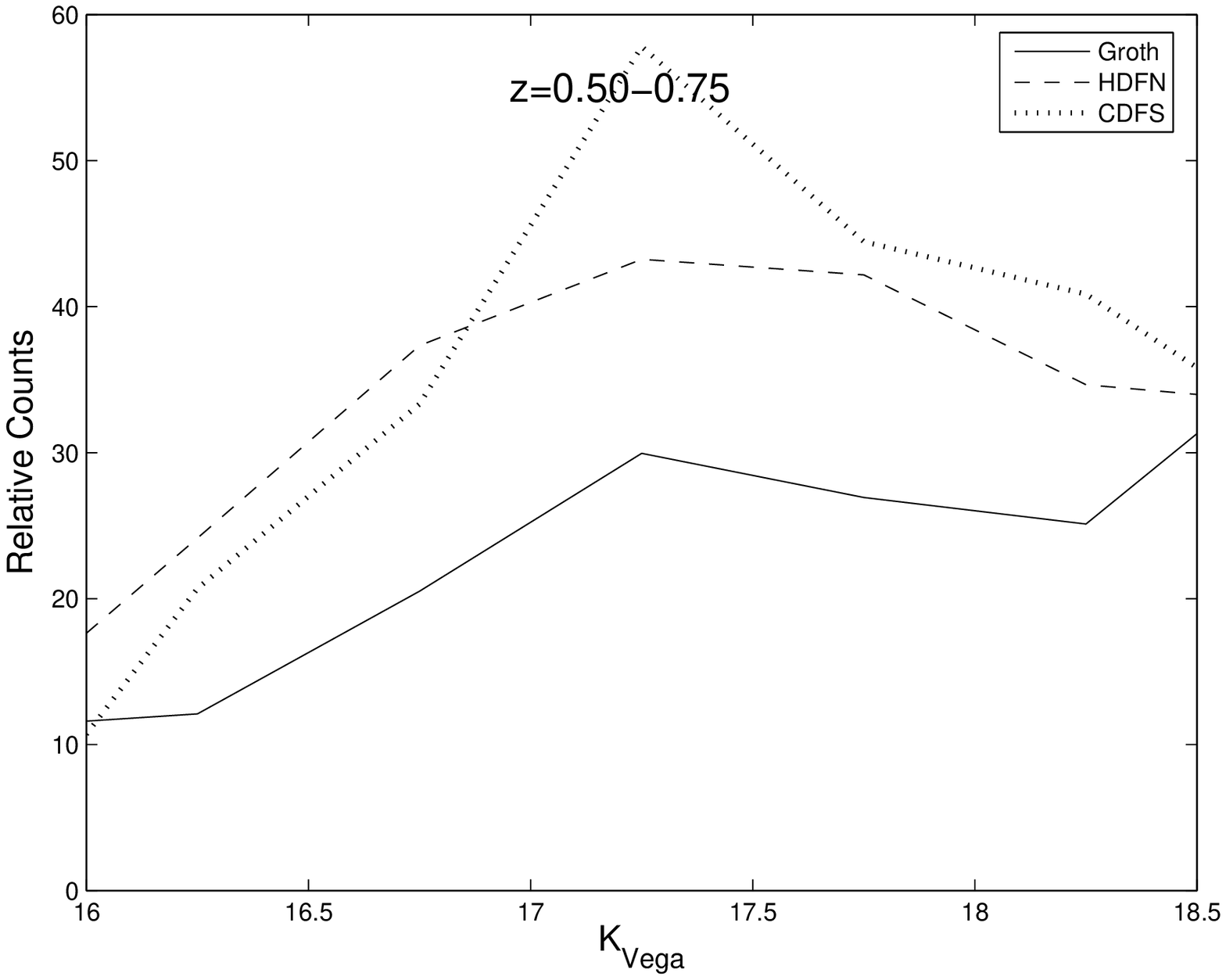}}\\
\resizebox{0.38\hsize}{!}{\includegraphics{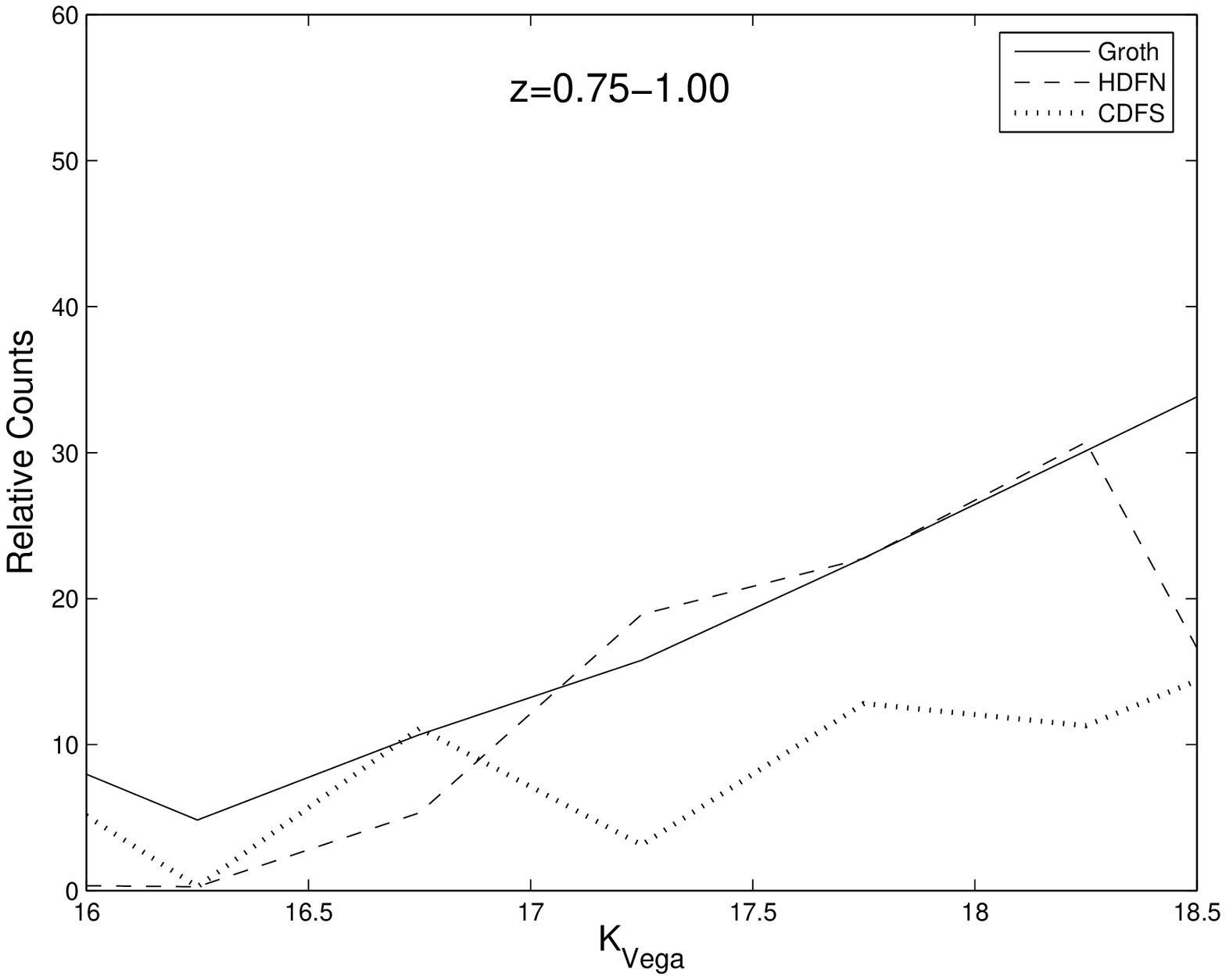}}
\resizebox{0.38\hsize}{!}{\includegraphics{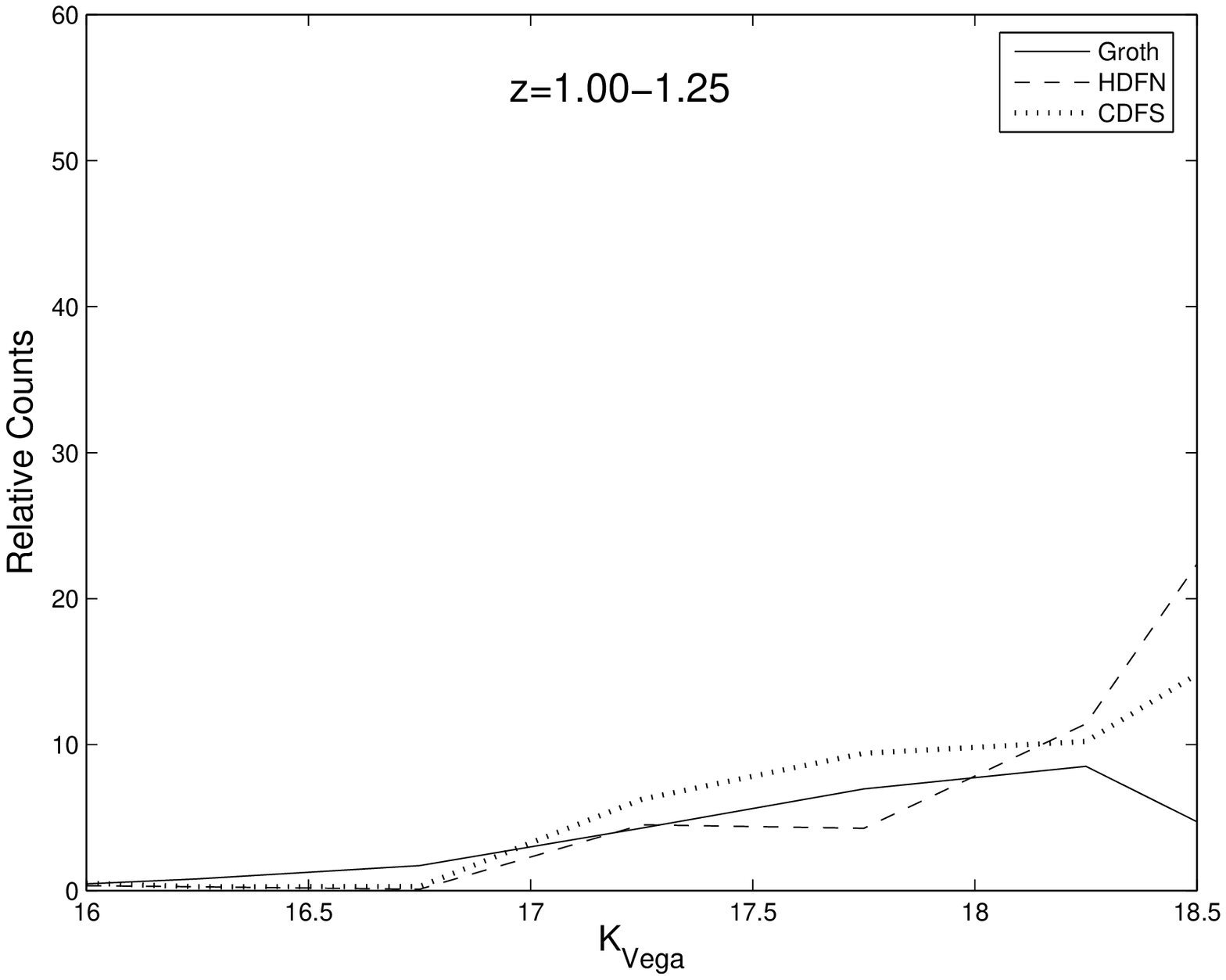}}
\caption{\label{counts_binned} Relative contribution to the
$K$-band number counts in redshift bins for the samples in the
HFDN, CDFS and Groth field. Field-to-field differences can reach
the $40\%$ for the most prominent LSS features.}
\end{figure*}

The four panels of Fig.~\ref{counts_binned} show the relative contribution to NCs in HDFN, CDFS and the Groth field divided in four  redshift bins: [0.25-0.50],[0.50-0.75],[0.75-1] and [1-1.25]. Below $z<0.25$ the averaged contribution is small enough ($<10\%$) to be neglected. The next three bins are responsible for most of the total counts up to $K\sim18$. However, the exact proportion might differ by up to a factor of 2 due to LSS. The alternate peaks and valleys between HDFN/CDFS and the Groth fields between $z=0.4-0.9$ (see Fig.~\ref{zphot_distrib}) cause the most prominent differences around $K=17.0-17.5$ in the first panel of Fig.~\ref{counts_binned}, similarly to the underdensity in CDFS at $z\sim$0.8 which leads to the significantly lower NCs around $K\sim17.5-18$ in the third panel.

\section{Number counts \& luminosity functions}\label{RESULTS}

In the previous section, we showed evidence for a mild decreasing trend in the slope of the NCs. In this section, we will try to find the origin of such a decreasing trend by studying the LF at several redshift intervals (benefiting from the estimation of photometric redshifts for the entire sample). The calculation of LFs must take into account that the observed $K$-band probes progressively bluer rest-frame bands with increasing redshift. By $z\sim0.75$, the $K$-band central wavelength shifts to $1.25\mu m$($J$-band), and for $z>1.5$ it begins to probe the optical bands. Therefore, in order to derive NCs from LFs it is necessary to select different rest-frame bands at each redshift bin.

In $\S$\ref{RESULTS}.1, we present the functional relation between the NCs and LFs,  parametrized using a \cite{1976ApJ...203..297S} function, that we will use to derive the $K$-band NCs. In $\S$\ref{RESULTS}.2, we derive LFs from the observed $K$-band, in the redshift range [0.25-1.25], using our three galaxies samples. In $\S$\ref{RESULTS}.3 we combine our LFs with LFs from the literature to explore the general picture of luminosity evolution in the optical and NIR. Finally, in $\S$\ref{RESULTS}.4, we summarize the multi-wavelength LFs probed by the observed $K$-band at different redshifts, that gives rise to the $K$-band NCs distribution.

\subsection{Number counts from Schechter functions}\label{MODEL}

The distribution of observed galaxy counts is a consequence of the LFs and the cosmological framework. Hence, any feature in the NC distribution could be explained in terms of the evolution of the LFs assuming a cosmological context. Here we will adopt a $\Lambda$CDM framework, and assume that the galaxy LF can be described by means of a Schechter function,

\begin{equation}
\label{schechter}
\phi(M)=0.4\ln(10)\phi^{\ast}10^{0.4(M^{\ast}-M)(\alpha+1)}\exp\big(10^{0.4(M^{\ast}-M)}\big)
\end{equation}
where $M^{\ast}$ is the characteristic absolute magnitude, $\alpha$ the faint-end slope, and $\phi^{\ast}$ the normalization of the luminosity function.
Although a Schechter function might not be appropriate to fully describe the faint end population of the LF \citep{2005ApJ...631..208B}, this is not the case for the bright NCs, which will be mostly dominated by $M^{\ast}$ galaxies (see next section).
The NCs represent the distribution of galaxies per apparent magnitude and sky area. Assuming a parametrization for the LF, the functional form of the NCs becomes

\begin{equation}\label{schechter_int}
\mathbf{N}(m_{\lambda_{0}},[M,\phi,\alpha]_{\frac{\lambda_{0}}{(1+z)}})=\int^{z_{f}}_{z_{i}}\phi(m,z,[M,\phi,\alpha]_{\frac{\lambda_{0}}{(1+z)}})\frac{dV_{\mathrm{c}}}{d\Omega}dz \label{eq_schechter}
\end{equation}
where $z_{i}$ and $z_{f}$ are the lower and upper limit of the redshift bin, $dV_{\mathrm{c}}/d\Omega$ is the differential comoving volume, and $m[M,\phi,\alpha]_{\frac{\lambda_{0}}{(1+z)}}$ represents the Schechter parameters of a LF in a band with effective wavelength $\frac{\lambda_{0}}{(1+z)}$. This indicates the explicit dependence of the NCs in the $K$-band on bluer LFs with increasing redshifts.
Note that even if we assume that the Schechter parameters are constant in the redshift bin $(z_{i},z_{f}]$, the expression \ref{schechter_int} still depends on LFs at different wavelengths. 
When $\Delta z$ is small, we can approximate the NCs from single LFs in a photometric band at $\frac{\lambda_{0}}{(1+\bar{z})}$, where $\bar{z}$ is the mean value of the redshift bin $(z_{i},z_{f}]$.

Additionally, Eq.~\ref{schechter_int} can be interpreted as a sum of LFs at different redshift bins, weighted by the corresponding comoving volumes. Hence, the shape of the NCs would be the result of the LF parameters at a given epoch and their evolution with redshifts modulated by the volume element.
Furthermore, the slope of the total NCs is the sum of the slope from each redshift bin weighted by the normalized NCs.

\begin{eqnarray}
\ln\Big(\frac{dN}{dm d\Omega}\Big)=\frac{1}{N}\frac{dN}{dm}=\sum_{i=0}^{\infty}\frac{\mathbf{N_{i}}}{{N}}\Big(\frac{1}{\mathbf{N_{i}}}\frac{d\mathbf{N_{i}}}{dm}\Big) \label{THEO_SLOPE}
\end{eqnarray}
where $\mathbf{N_{i}}$ are the NCs in the redshift bin $i$, and $N$ is the sum of all the redshift bins.
The slope of the NCs presents some interesting properties. First, it is independent of the absolute value of $\phi^{\ast}$, but it does depend on the relative change of this parameter (i.e., it depends only on the evolution of $\phi^{\ast}$), which can be normalized arbitrarily. Second, in addition to the low redshift Euclidean limit ($d\log N/dm=$0.6), it can be shown that, at very faint magnitudes, the slope approximates asymptotically to $-0.4(\alpha+1)$.
\begin{equation}
\ln\Big(\frac{dN}{dm d\Omega}\Big)\Big\arrowvert_{m\gg}=0.4\Big(\frac{I(\alpha,M(m))}{\hat{I}(\alpha,M(m))}10^{-0.4m}-(\alpha+1)\Big)\label{slope_limit}
\end{equation}
where $I(\alpha,M(m))$ accounts for all other terms except $10^{-0.4m}$, including the integral over redshift and the exponential term in the Schechter equation, and $M(m)$ indicates the relation between absolute and apparent magnitudes through the luminosity distance. The $I(\alpha,M(m))$ in the numerator differs from the denominator only in the derivative of the exponential term, which also outputs the $10^{-0.4m}$ term, indicated explicitly. 
At faint magnitudes the first term of the Eq.~\ref{slope_limit} tends to zero, dominated by the faster decrement of $10^{-0.4m}$,  yielding the asymptotic limit $-0.4(\alpha+1)$. More intuitively, it can be understood as a regime where the NCs are mostly dominated by the faint end of a single LF. Hence, the slope of the NCs necessarily becomes the slope of this LF. For any given variation with redshift of the Schechter parameters, there will always be a maximum value of the product $\phi^{\ast}\frac{dV_{c}}{d\Omega}$. The LF at that redshift will be favored among the others (suppressed smaller values of the product) and, at sufficiently faint magnitudes, it will be the main contributor to the NCs.

A very interesting result that emerges from the analysis of the NCs in the full magnitude range is that the evolution of the slope can be summarized in three main regimes, namely: 
\begin{itemize}
\item (1) {\it The Euclidean regime}. The classical low redshift approximation, that yields the well-known result $d\log N/dm=0.6$. The NCs in this regime are mostly populated by M$\leq M^{\ast}$ at very low redshift (z$<0.2$).
\item (2) {\it The transition regime}. The slope departs from the Euclidean limit into a weighted sum of the slopes of LFs at low-mid redshifts. The dominant contribution would come from the LF at the redshift that maximizes the product $\phi^{\ast}\frac{dV_{c}}{d\Omega}$ (the factors controlling the normalization of Eq.~\ref{schechter_int}). In this regime, the slope decreases rapidly around the apparent magnitude of $M^{\ast}$ at the dominant LF (i.e. at the knee of the LF). In the absence of a significant evolution in $\phi^{\ast}$, the maximum would take place at the peak of the volume element at z$\sim$2 \citep{2003RMxAC..16..203B}.
\item (3) {\it The ``$\alpha$ regime''}. In this regime, the majority of the LFs dominating the NCs contribute with galaxies at the faint end. Therefore, the slope will be a combination of individual slopes approaching the minimum value ($\sim$-0.4($\alpha$+1)). As a consequence of the previous regime, this phase should be controlled by the same LF as in the previous phase. However, other LFs, not strongly suppressed by the $\phi^{\ast}\frac{dV_{c}}{d\Omega}$ factor, and having a significantly larger $\alpha$, might dominate this phase.
\end{itemize}
\noindent
We will further discuss the evolution of the slope in terms of this schema in $\S$\ref{NCs from LFs}.1. 

\subsection{LFs from the observed K-band}
\begin{figure*}
\centering
\resizebox{0.4\hsize}{!}{\includegraphics{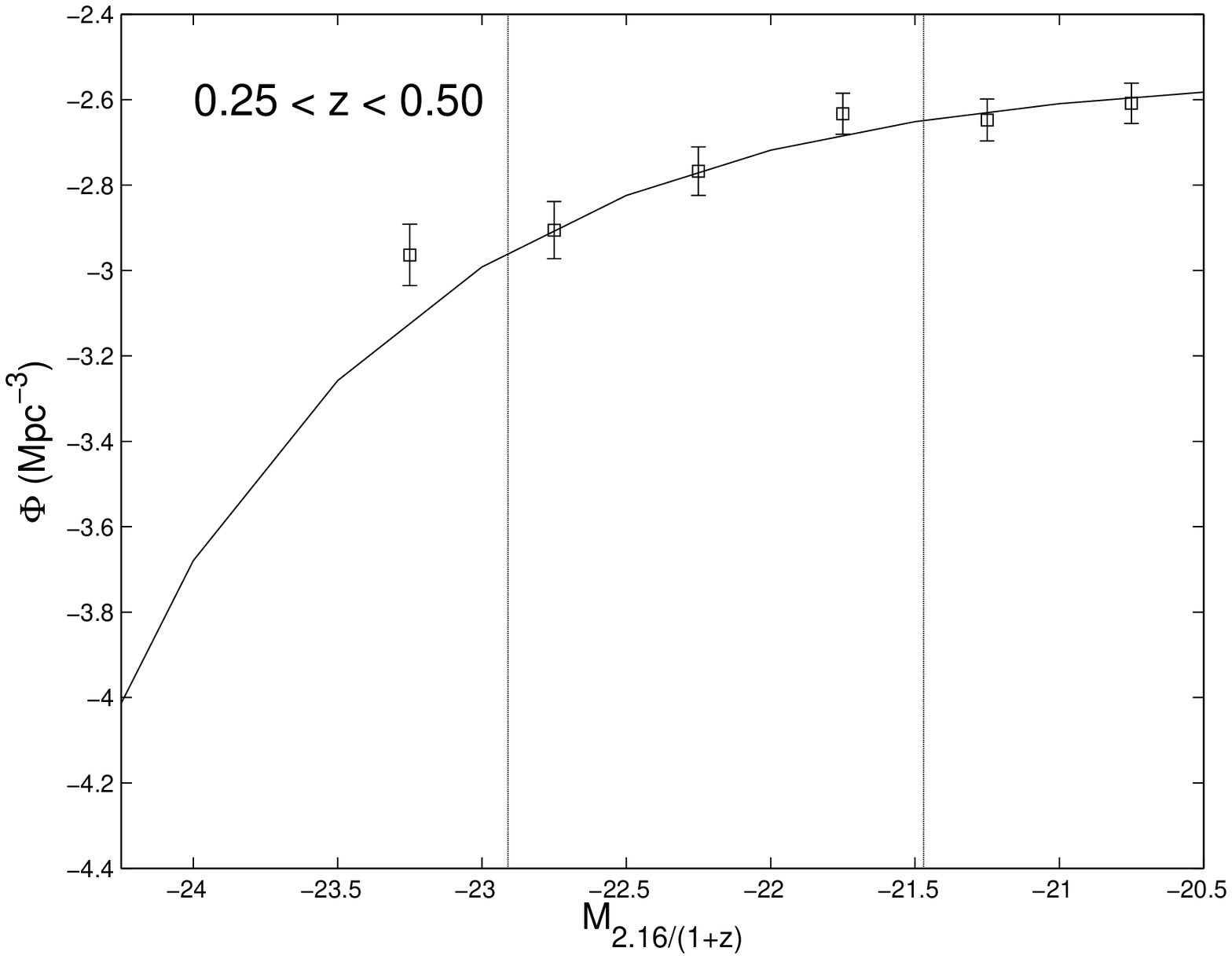}}
\resizebox{0.4\hsize}{!}{\includegraphics{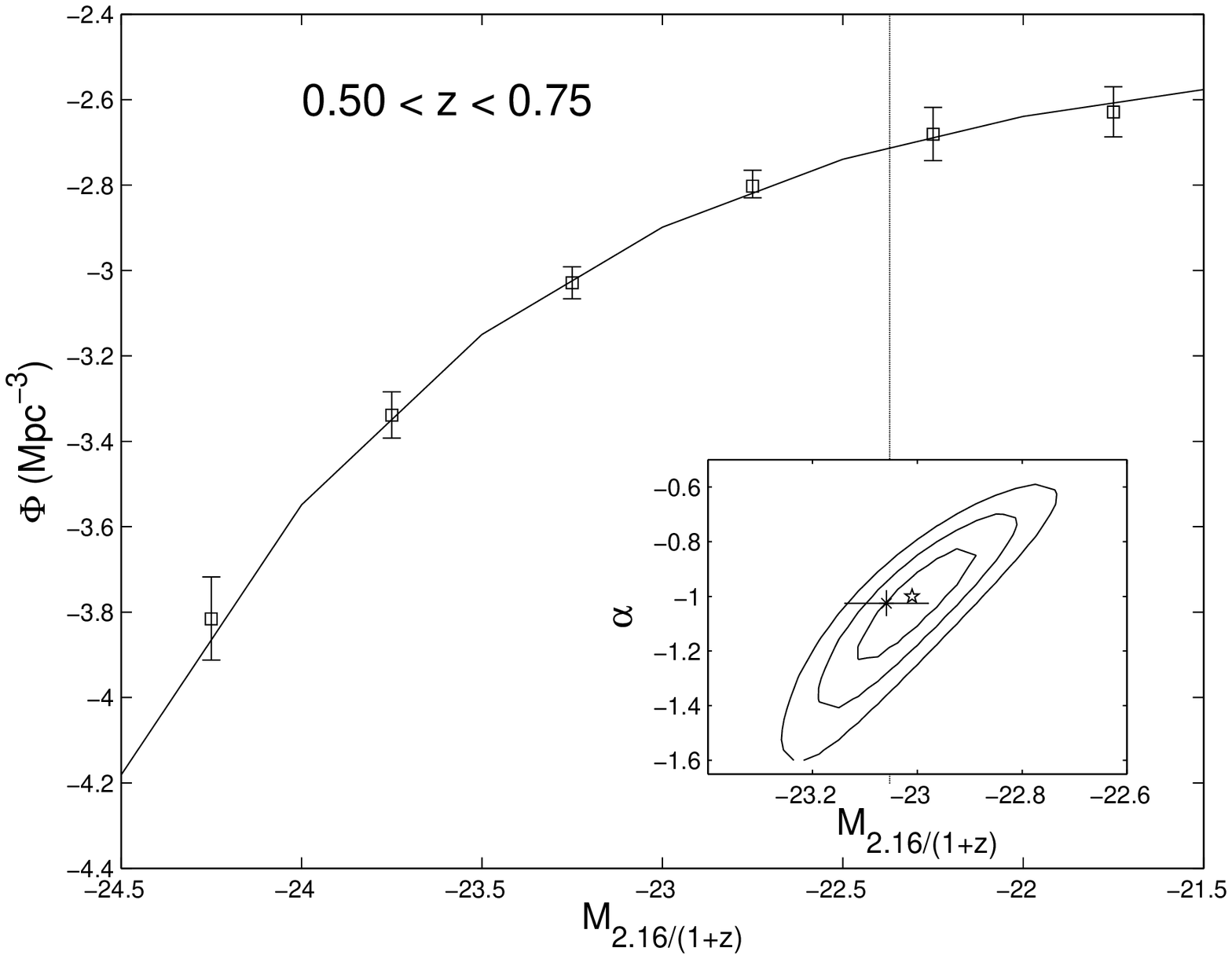}}\\
\resizebox{0.4\hsize}{!}{\includegraphics{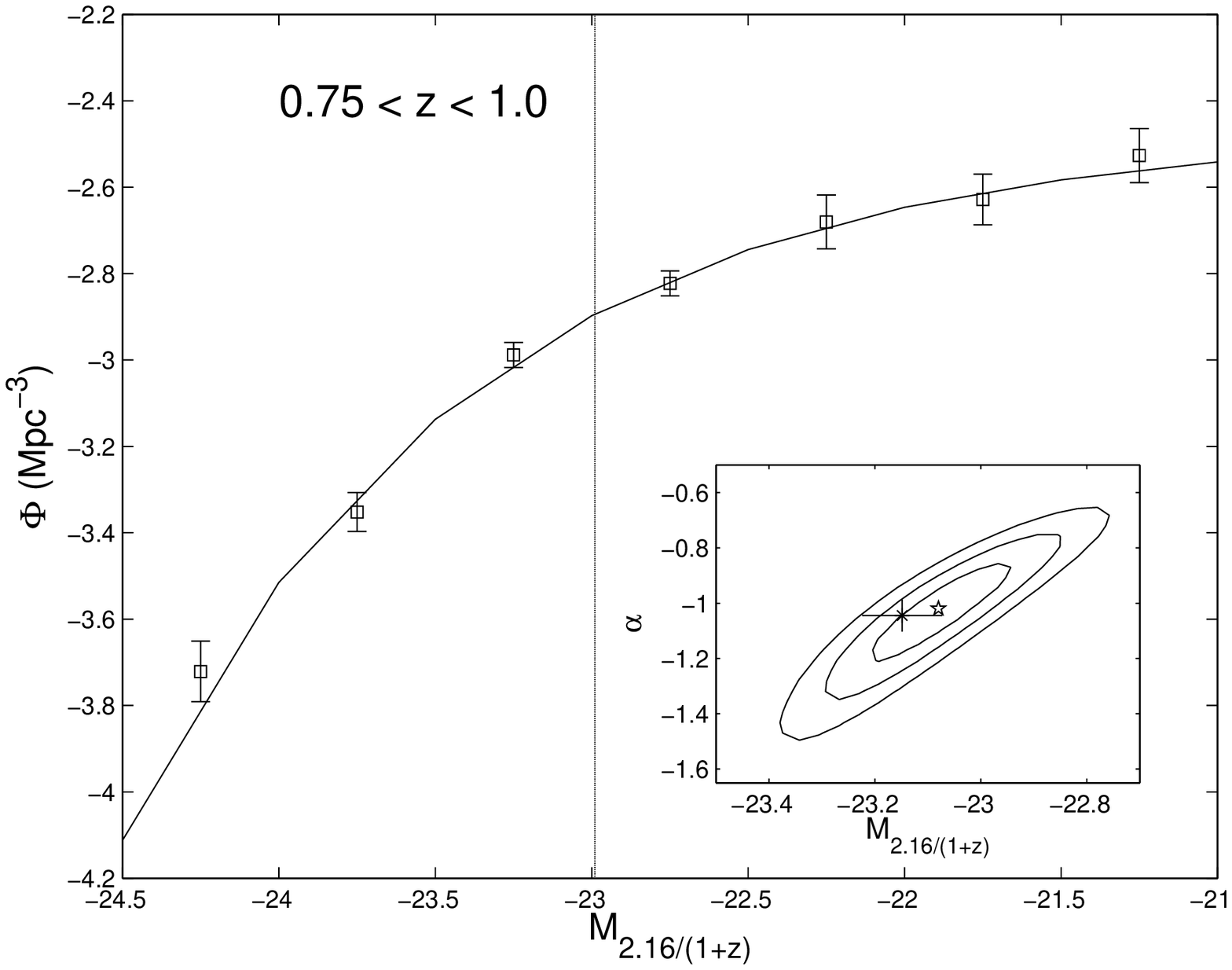}}
\resizebox{0.4\hsize}{!}{\includegraphics{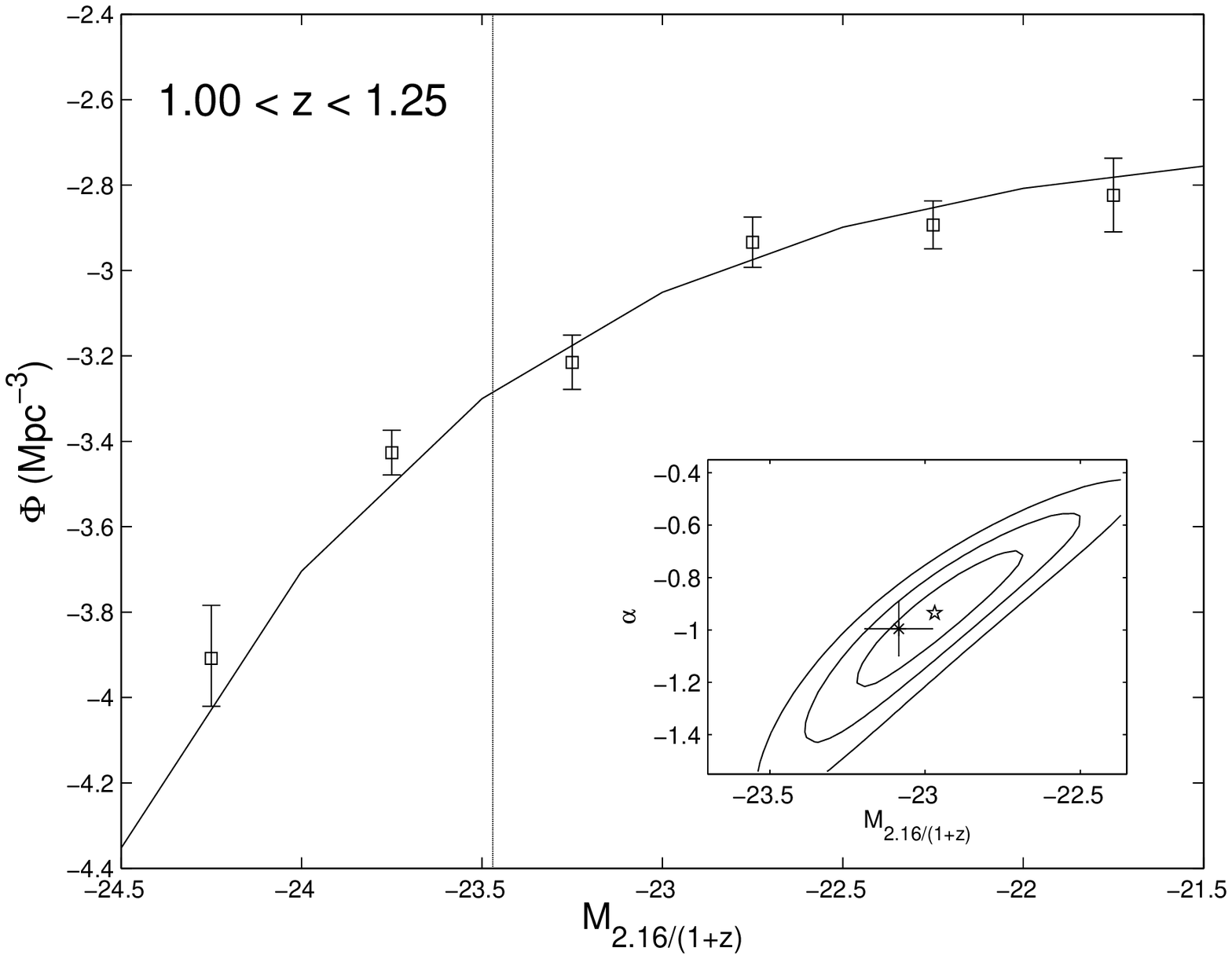}}
\caption{\label{LF_plot} LFs derived from the observed $K$-band in different redshift slices. The LF with the $V_{\mathrm{max}}$ method are shown as open squares with error bars (including Poisson, photometric and photometric redshift uncertainties).The continuous line depicts the best fit from the STY method. The inset shows the best fitting ($M$,$\alpha$) values (star) with 1,2 and 3$\sigma$ confidence contours, and also the mean value (cross) with error bars derived from the photometric redshift simulations. The vertical line shows the photometric threshold for the shallowest fields, $K_{\mathrm{Vega}}=18.5$. The additional vertical line in the first panel shows the lower threshold at bright magnitudes.}
\end{figure*}

In order to derive NCs in the $K$-band using Eq.~\ref{schechter_int} we need to feed the equation with LFs in different bands at different redshifts. More precisely, we are interested in the LFs from the observed $K$-band. Thus, we have obtained LFs in several redshift bins at decreasing rest-frame wavelengths centered at $2.16\mu m/(1+\bar{z})$, where $\bar{z}$ is the mean redshift of the bin $(z_{i},z_{f}]$.

\subsubsection{Methodology}
To estimate the observed LF we have applied the $1/V_{\mathrm{max}}$ method \citep{1968ApJ...151..393S}. We associate the Poisson errors to each magnitude bin \citep{1986ApJ...303..336G} added in quadrature to photometric errors. Since we have not applied k-corrections, $\Delta m$ translates directly into $\Delta M$. Additionally, in order to quantify the uncertainties due to the photometric redshift errors, we have performed Monte Carlo simulations at each redshift bin, assuming an uncertainty characterized by the width of a Gaussian distribution derived fitting the histogram of $\sigma_{z}/(1+z)$ for each redshift bin.

Additionally, we measured the LF from the STY \citep{1979ApJ...232..352S} estimator, which is a parametric maximum likelihood method. The STY is not as sensitive to large-scale fluctuations as the $1/V_{\mathrm{max}}$ method, but requires a priori fixing the functional form of the LF. Nevertheless, the assumption that $\phi(M)$ is well described by a Schechter function suits our purpose of explaining the NCs in terms of LFs, as we have described in the previous section.

To check the effect of photometric redshift errors in the STY method, we perform the same Monte Carlo simulations as for the $1/V_{\mathrm{max}}$. Since the probability of galaxies existing at redshift $z$ grows rapidly until $z\sim2.5$, a larger fraction of intrinsically faint sources are shifted to the bright end than viceversa. This can potentially bias the determination of the LF introducing large systematic effects when $\Delta z$ is large and the distance modulus grows rapidly (\citealt{2003ApJ...586..745C}; \citealt{2007ApJ...656...42M}). Nevertheless, the combination of the typical values of $\sigma_{z}\sim0.05-0.07$ with the relatively shallow limiting magnitude of our LFs (less than $\sim1$ mag deeper than $M^{\ast}$) cause only a moderate offset in the Schechter parameters ($\Delta M < 0.2$,$\Delta\alpha < 0.1$). It can be seen from the inset of Fig.~\ref{LF_plot} that the mean and standard deviation resulting from the simulations (star) lie always within the 1$\sigma$-2$\sigma$ confidence level derived from the best fitting result of the STY method.

\subsubsection{Observed luminosity functions}
The LFs, $\phi(M)dM$, of the combined samples were computed in four redshift bins of width $\Delta z=0.5$ using the methods described above.

\begin{table}[h]
\centering
\small
\begin{tabular}{cccc}
\hline
redshift bin & $M^{\ast}_{K,obs}$(AB)     & $\alpha_{K,obs}$              & $\phi^{\ast}_{K,obs}$(10$^{-3}$Mpc$^{-3}$) \\
\hline
0.25 - 0.50       & $-22.95$(fixed) & $-1.15^{0.20}_{0.20}$ &        $3.4^{2.10}_{1.70}$       \\
0.50 - 0.75       & $-23.01^{0.13}_{0.12}$ & $-1.00^{0.18}_{0.23}$ &        $3.4^{0.22}_{0.18}$       \\
0.75 - 1.00       & $-23.08^{0.14}_{0.12}$ & $-1.02^{0.19}_{0.17}$ &        $3.2^{0.17}_{0.24}$       \\
1.00 - 1.25       & $-22.96^{0.28}_{0.26}$ & $-0.93^{0.23}_{0.29}$ &        $2.3^{0.31}_{0.26}$       \\
\hline
\end{tabular}
\caption{\label{LF_summary} Best fitting Schechter parameters for the STY LFs from our combined samples. The 1$\sigma$ errors were derived from Monte Carlo simulations}
\end{table}

Table~\ref{LF_summary} summarizes the best fitting values of the Schechter function for each redshift. Fig.~\ref{LF_plot} shows the estimates from the $V_{\mathrm{max}}$ method (open squares) with $1\sigma$ error bars (from Poisson statistics and photometric redshift simulations), and the LF from the maximum likelihood method at different redshifts. The inset shows the 1, 2 and 3$\sigma$ confidence levels for the $\alpha$ and $M^{\ast}$ parameters (open star), together with the mean value derived from the photometric redshift simulations (cross). The vertical line indicates the photometric threshold of the shallowest sample ($K$=18.5). For fainter luminosities, the values of the LFs are derived exclusively from the CDFS sample. Note that the completeness limit of our samples translates into luminosities very close to or even fainter than $M^{\ast}$ for the higher redshift bins. This means that galaxies fainter than $M^{\ast}$ will not contribute significantly to the NCs up to $K<18.5$. Nevertheless, the deeper coverage of the CDFS allow us to properly fit the LFs that will be employed to derive the NC function (see section \S\ref{LF_from_Kband}).

In order to properly account for cosmic variance, the values of $\phi_{i}$ for the $V_{\mathrm{max}}$ method derived exclusively from the CDFS data (those above the photometric threshold; vertical line of Fig.~\ref{LF_plot}) have been corrected by a scale factor. This scale factor has been derived from the median value of all fields in the magnitude bins in common, weighted by the area covered by each field. Also, only the magnitude bins with more than 100 galaxies in the CDFS sample have been considered in the calculation.

Finally, for the $0.25<z<0.50$ bin, the bright extremity of the LF is poorly constrained even in the total combined area of the three fields. Nevertheless, since the faint end is properly sampled, we decided to fit the LF fixing the value of $M^{\ast}$ and setting a threshold on the bright ($K<16$) magnitudes. The value of $M^{\ast}$ was taken from the literature. At $z\sim0.38$ the rest-frame wavelength probed by the $K$-band is close to the $H$-band. However, no references for $H$-band LFs at that redshift were available then, and we decided to use $M^{\ast}$ from the rest-frame $K$-band LF at $z\sim0.4$ published by \cite{2007arXiv0705.2438A}.Then, we corrected that value by applying an $H$-$K$ color term. The mean value of that color at $z\sim0.4$ derived from our data is $\langle H-K\rangle(AB)=-0.21\pm0.11$, very similar to the local value ($\sim$-0.20; \citealt{2003AJ....125..525J}).

\begin{figure*}\centering
\resizebox{!}{0.35\hsize}{\includegraphics{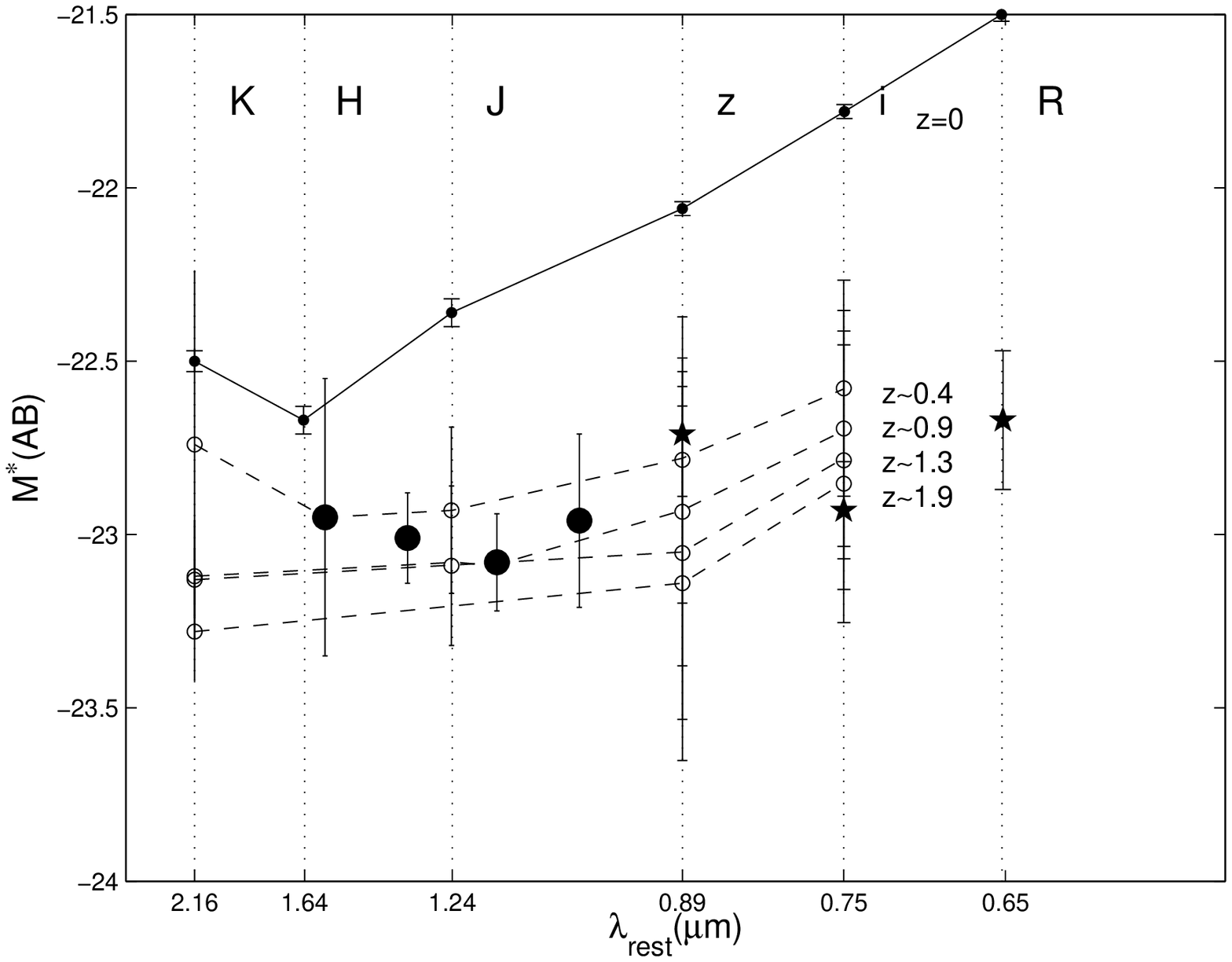}}
\resizebox{!}{0.35\hsize}{\includegraphics{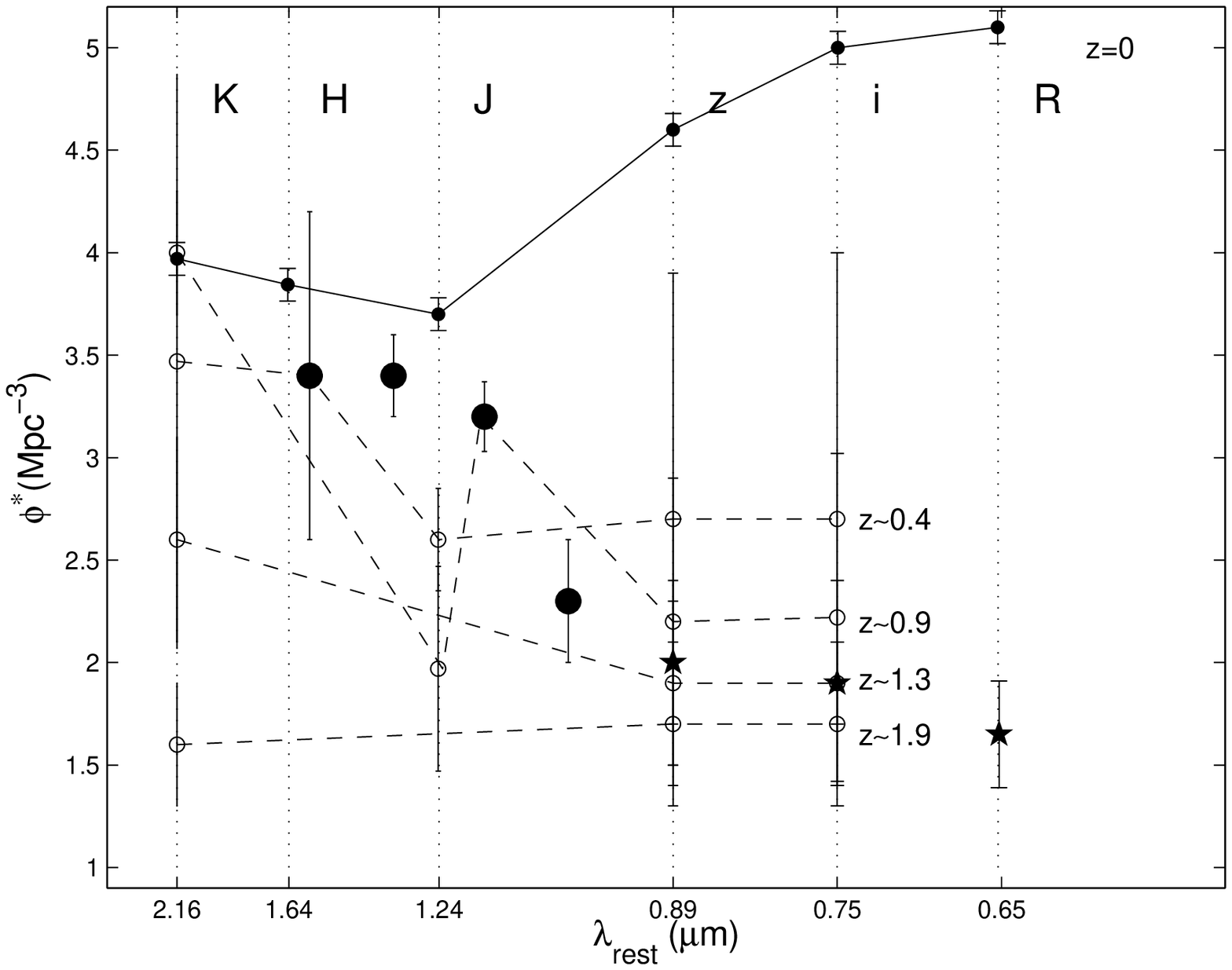}}
\caption{\label{evolving_param}Evolution with redshift of the
Schechter parameters $M^{\ast}$(left panel), $\phi^{\ast}$(right panel) in the $RizJK$ rest-frame bands. For
each photometric band (RizJK, vertical dotted lines) the empty circles depict the redshift evolution of the parameter at $z\sim$ 0.4, 0.9, 1.3 and 1.9, drawn from different authors (see Table~\ref{LF_summary}). The long dashed lines connect the values of the parameter at the same redshift in different bands.
The black stars shows the value of the parameter at $z=1.70$, $z=2.26$ and $z=2.25$ in the i, z and R bands respectively. The black dots indicate the LF parameters at $\bar{z}=0.38, 0.62, 0.88, 1.12$ derived in this work.}
\end{figure*}

\begin{figure*}\centering
\resizebox{!}{0.35\hsize}{\includegraphics{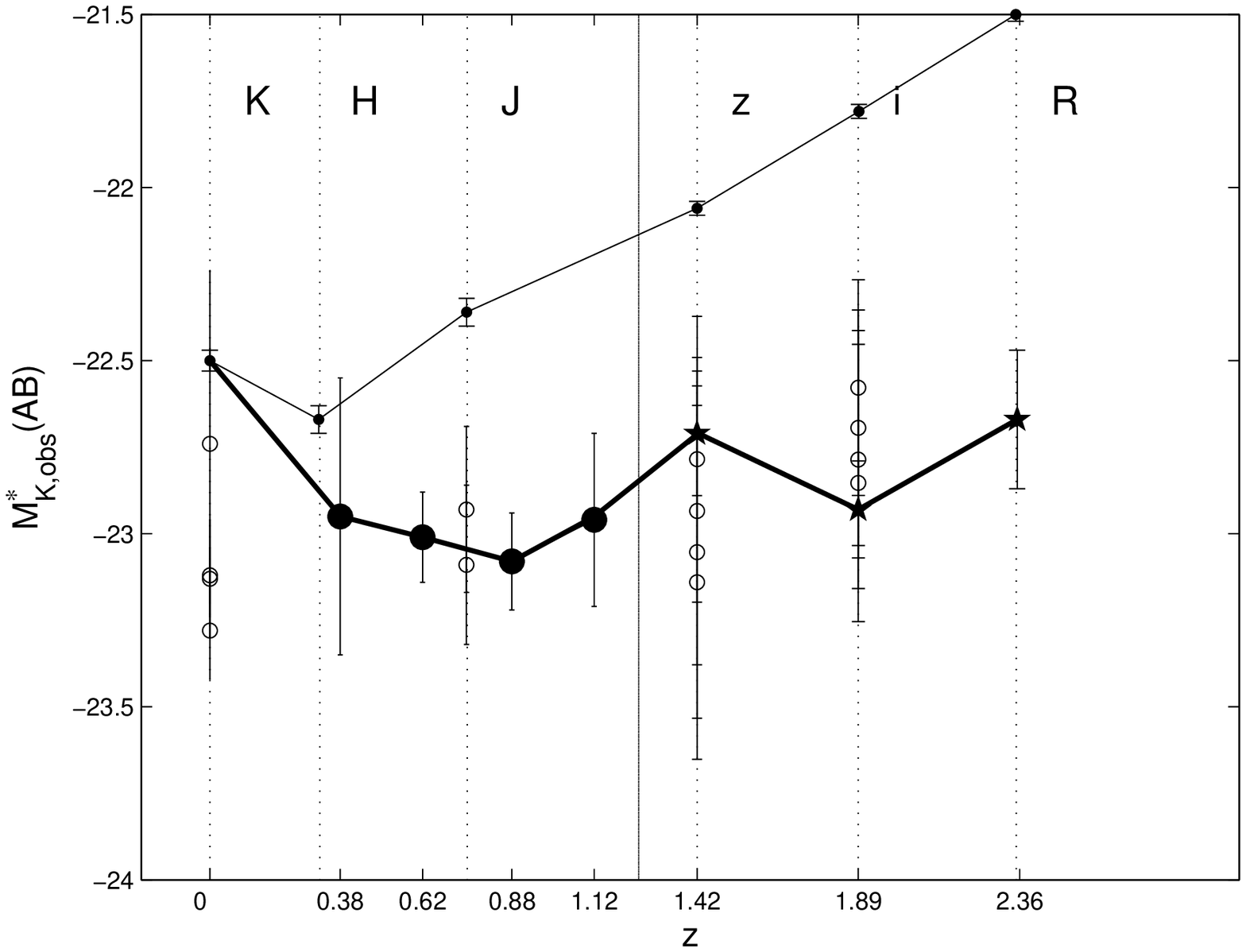}}
\resizebox{!}{0.35\hsize}{\includegraphics{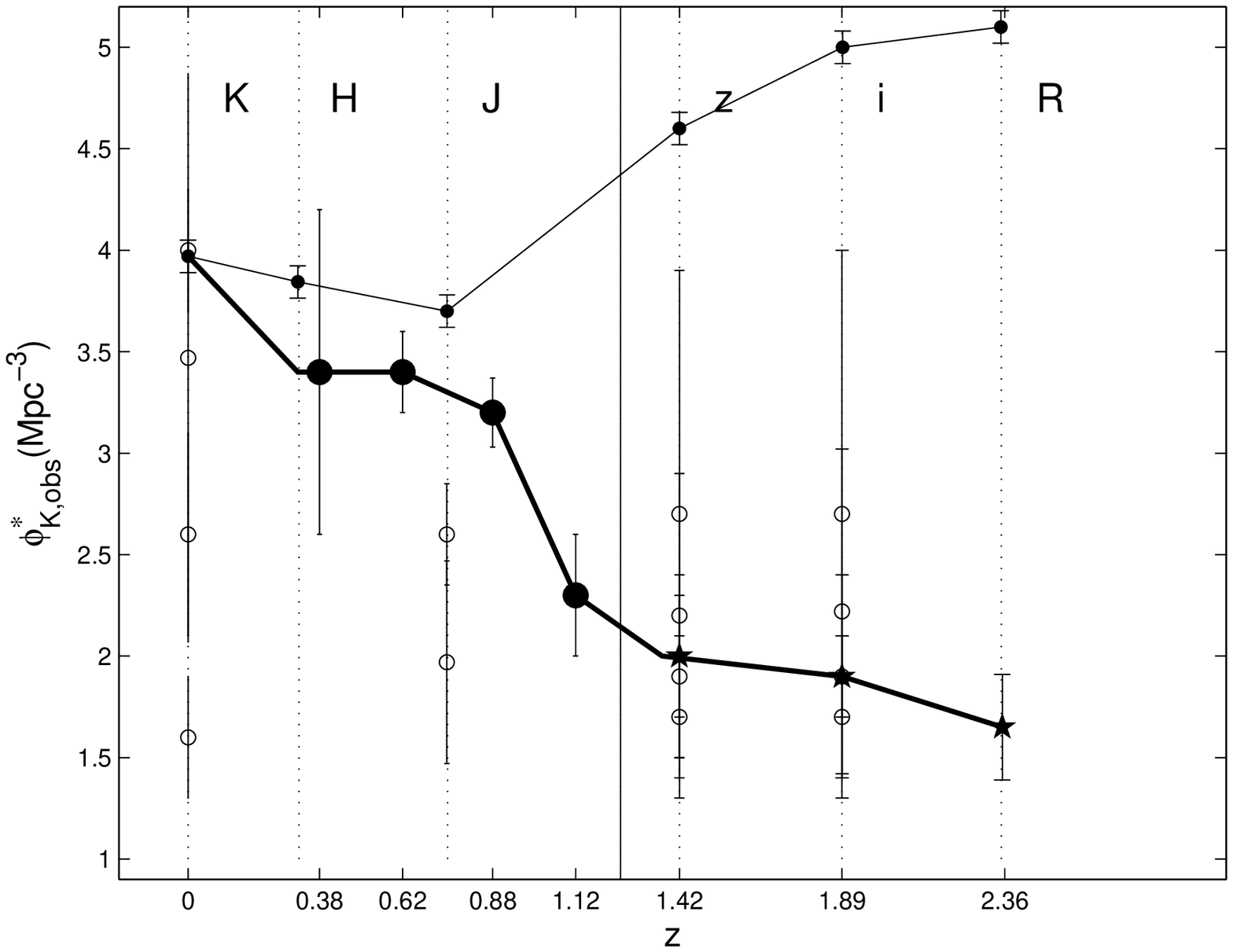}}
\caption{\label{evolving_param_Kobs}Multi-wavelength evolution of the
Schechter parameters $M^{\ast}_{K,obs}$ (left panel), $\phi^{\ast}_{K,obs}$ (right panel) as probed by the observed $K$-band. For each photometric band (RizJK, vertical dotted lines) the empty circles depict the
redshift evolution of the parameter at $z\sim$ 0.4, 0.9, 1.3 and 1.9, drawn from different authors (see Table~\ref{LF_summary}). The black stars show the value of the parameter at $z=1.70$, $z=2.26$ and $z=2.25$ in the i, z and R bands respectively. The grey vertical line at $z=1.25$ is the estimated redshift from which the contribution of the sources to the NCs up
to $K=18.5$ is less than $10\%$. The black dots indicate the LF parameters at $\bar{z}=0.38, 0.62, 0.88, 1.12$ derived in this work. The thick black line shows the approximated evolution of the Schechter parameters as probed by the observed $K$-band at growing redshifts.}
\end{figure*}

\subsection{LFs evolution in rest-frame bands}

Analyzing the evolution of the LFs directly derived from the observed $K$-band requires exploring LFs in shorter rest-frame wavelengths at different redshifts. For this purpose we compiled LF estimates from the literature in optical and NIR bands. We have compiled the local LFs in the $J$ and $K$-bands from \cite{2001MNRAS.326..255C} and \cite{2001ApJ...560..566K} respectively, whereas for the $H$-band we estimated a value of $M^{\ast}$ by applying a color correction to the $K$-band value, following the same procedure as in the previous section. The local optical LFs in the i,z and R bands were drawn from \cite{2003ApJ...592..819B}.

At higher redshift, the $K$-band LFs were drawn from \cite{2007arXiv0705.2438A}.
Their LFs for the {\sc swire-vvds-cfhtls} survey predicts an evolution in the Schechter parameters
which is consistent with previous results from \cite{2007MNRAS.380..585C} in the {\sc ukidds-uds}. For the $J$-band we considered the LF estimates from \cite{2003A&A...402..837P} (K20 survey, 52~arcmin$^{2}$), \cite{2003MNRAS.342..605F} (MUNICS survey, 0.17~deg$^{2}$), and \cite{2005ApJ...631..126D} (GOODS-CDFS, 130~arcmin$^{2}$). Their results are compared in Figure 15 of the latter, showing that, despite the
apparent inconsistency in the Schechter parameters, there is good agreement in the data points. The different observational constraints of each survey lead to different best fitting values for a distribution with a very similar shape. We have taken the LF in the~ $J$-band at $z\sim0.48$ from \cite{2003MNRAS.342..605F} and from \cite{2005ApJ...631..126D} at $z\sim0.9$. The larger area of the first is better to constraint the bright end at low redshift, while the depth of the latter is more suitable at higher redshifts.

We find the same apparent inconsistencies due to the $\alpha-M^{\ast}$ degeneracy in the LFs of optical bands. \cite{2005A&A...439..863I} (VVDS, 0.6~deg$^2$) and \cite{2006A&A...448..101G} (FORS Deep Field Survey, FDF, 35~armin$^2$) derived very different values of the Schechter parameters that nevertheless produce $1-2\sigma$ compatible LFs when comparing both estimates under a common limiting magnitude and using Monte Carlo simulations \citep{2006A&A...448..101G}. For our purposes we are more interested in the high-redshift optical LFs ($z>1.5$). Thus, we preferred the estimates from \cite{2006A&A...448..101G} that, despite the smaller area, are  able to probe the LF 3-4 magnitudes deeper. In addition to the LFs at $z=1.60$ and $z=2.26$, we derived some values for $M^{\ast}$ and $\phi^{\ast}$ at other redshifts using the evolutionary parametrization  also given in Gabasch et al. (2006; as $(1+z)^{\gamma}$).
Finally, the LF in the R-band at $z\sim2.3$ was drawn from \cite{2007ApJ...656...42M}. The authors derive high redshift LFs in optical bands from a combination of very deep nIR infrared observations (including 51arcmin$^{2}$ to $K\sim23$). Table~\ref{LF_summary_authors} summarizes the Schechter parameters of the LFs in the different bands and redshifts.

\begin{table*}[t]
\scriptsize
\begin{tabular}{cccccccc}
\hline
Source & Limit & Area & Band & z & $M^{\ast}-5\log h_{70}$ & $\phi^{\ast}$ & $\alpha$  \\
       & [mag] & [$\mathbf{deg}^{2}$] & & & [AB] & [$10^{-3}h_{70}^{3}Mpc^{-3}$] & \\
\hline
Kochanek et al. (2001) & K$< 11.25$           &$<7000$  & K &    0.02     & $-22.36\pm0.05$ & $3.97\pm0.34$ & $-1.09\pm0.06$ \\
Arnouts  et al. (2007) & F(3.6$\mu m)>9\mu Jy$& 0.85 & K &    0.50     & $-22.83\pm0.30$ & $3.47\pm1.4$  & $-1.1\pm0.2$   \\
Arnouts  et al. (2007) & F(3.6$\mu m)>9\mu Jy$& 0.85 & K &    0.90     & $-23.12\pm0.08$ & $4.00\pm0.3$  & $-1.11\pm0.06$ \\
Arnouts  et al. (2007) & F(3.6$\mu m)>9\mu Jy$& 0.85 & K &    1.35     & $-23.13\pm0.17$ & $2.60\pm0.5$  & $-1.1\pm0.2$   \\
Arnouts  et al. (2007) & F(3.6$\mu m)>9\mu Jy$& 0.85 & K &    1.75     & $-23.28\pm0.14$ & $1.62\pm0.3$  & $-1.1\pm0.2$   \\
Feulner et al. (2003) &  K$< 17.50$ & 0.18 & J &  0.30..0.60 & $-22.93\pm0.24$ & $2.60\pm0.80$ & -1.00 fixed    \\
Dahlen et al. (2005)  &             &  0.036  & J &  0.75..1.00 & $-23.09\pm^{0.24}_{0.22}$ & $1.97\pm^{0.60}_{0.40}$ & $-1.31\pm^{0.10}_{0.09}$ \\
Gabasch at al. (2006) & I(AB)$< 26.8$ & 0.01  & z & 0.4 &  $-22.78\pm0.41$ & $2.7\pm1.2$ &$-1.33$(fixed) \\
Gabasch at al. (2006) & I(AB)$< 26.8$ & 0.01  & z & 0.9 &  $-22.93\pm0.44$ & $2.2\pm0.7$ &$-1.33$(fixed) \\
Gabasch at al. (2006) & I(AB)$< 26.8$ & 0.01  & z & 1.32 & $-23.05\pm0.48$ & $1.9\pm0.5$ &$-1.33$(fixed) \\
Gabasch at al. (2006) & I(AB)$< 26.8$ & 0.01  & z & 1.31..1.91 & $-22.71\pm0.18$ & $2.0\pm0.3$ &$-1.33$(fixed) \\
Gabasch at al. (2006) & I(AB)$< 26.8$ & 0.01  & z & 1.89 & $-23.14\pm0.51$ & $1.70\pm0.4$ &$-1.33$(fixed) \\

Gabasch at al. (2006) & I(AB)$< 26.8$ & 0.01  & i & 0.4 &  $-22.58\pm0.31$ & $2.7\pm1.3$ &$-1.33$(fixed) \\
Gabasch at al. (2006) & I(AB)$< 26.8$ & 0.01  & i & 0.9 &  $-22.69\pm0.34$ & $2.2\pm0.8$ &$-1.33$(fixed) \\
Gabasch at al. (2006) & I(AB)$< 26.8$ & 0.01  & i & 1.32 & $-23.78\pm0.37$ & $1.9\pm0.5$ &$-1.33$(fixed) \\
Gabasch at al. (2006) & I(AB)$< 26.8$ & 0.01  & i & 1.89 & $-22.85\pm0.40$ & $1.7\pm0.4$ &$-1.33$(fixed) \\
Gabasch at al. (2006) & I(AB)$< 26.8$ & 0.01  & i & 1.91..2.61 & $-22.93\pm0.14$ & $1.9\pm0.2$ &$-1.33$(fixed) \\

Marchesini et al. (2007) & $K<23$(max) & 0.65 & R & 2.0..2.5 & $-22.67\pm^{0.20}_{0.22}$ & $1.07\pm^{0.27}_{0.27}$ & $-1.01\pm^{0.21}_{0.20}$ \\
\hline
\end{tabular}
\caption{\label{LF_summary_authors}Summary of the derived parameters of the LF parameters obtained by various authors by fitting them with Schechter functions, in several rest-frame bands, in different redshift bins, and from different samples of field galaxies.}
\end{table*}

The two panels of Fig.~\ref{evolving_param} show the evolution
with redshift of $M^{\ast}$ and $\phi^{\ast}$ in several NIR and
optical bands. Vertical dotted lines indicate the effective
rest-frame wavelength of each band. The open circles
show the values of $M^{\ast}$ and $\phi^{\ast}$ in each band at
different redshifts increasing from top to bottom. For clarity, the local values have been plotted with small black dots. The dashed lines connect the values of the Schechter parameters for the different bands at the same redshift (continuous grey line at z=0). Finally, the large black dots show our estimates of the LFs at $z\sim0.4,0.6,0.9,1.1$. These estimates do not match exactly any standard photometric band. However, the value at $z\sim0.4$ is almost centered in the $H$-band, and the LFs at $z\sim0.6$ and $z\sim0.9$ should be consistent with the estimates in the $J$-band. Comparing our values with the results for these bands of Dahlen et al. and Feulner et al., we find a very a good agreement in the characteristic luminosity, whereas the values of $\phi^{\ast}$ are poorly consistent. A possible explanation for this discrepancy is that the values of Dahlen et al. at $z\sim0.9$ might be affected by an underdensity peak between $z=0.7-1.1$, while our results present a slight overdensity around that redshift (see Fig.~\ref{zphot_distrib}). Nevertheless, the comparison to the values of Feulner et al. is $2\sigma$ compatible, and is also consistent with the results of \cite{2003A&A...402..837P}. Wider areas are necessary to reduce the field-to-field variations and obtain more reliable estimates.

It can be seen from the left panel of Fig.~\ref{evolving_param} that the multi-band values of $M^{\ast}$ at $z=0$ resemble the shape of a galactic SED, with the absolute magnitude peaking around the H band, close to the $1.6\mu m$ stellar bump. At higher redshifts, the shape is not preserved due to the different luminosity evolution in each band; i.e., in the $K$-band, a substantial brightening of $\sim1~$mag
between the local universe and $z\sim2$ has been reported by several authors (\citealt{2006MNRAS.366..609C}; \citealt{2007MNRAS.380..585C}; \citealt{2006MNRAS.367..349S}). In the $J$-band, Saracco et al. find $\Delta M^{\ast}\sim-0.7$ to $z\sim3$, and the results from \cite{2007ApJ...656...42M} indicate a similar or slightly higher brightening ($\Delta M^{\ast}\sim1.7$) in the optical LFs up to the same epoch. However, it is not surprising that the brightening is faster in the optical bands than in the NIR since they are more affected by the light coming from younger stellar populations (\citealt{2005ApJ...631..126D}; \citealt{2005A&A...439..863I}; \citealt{2003A&A...401...73W}).

\begin{figure*}
\resizebox{!}{0.40\hsize}{\includegraphics{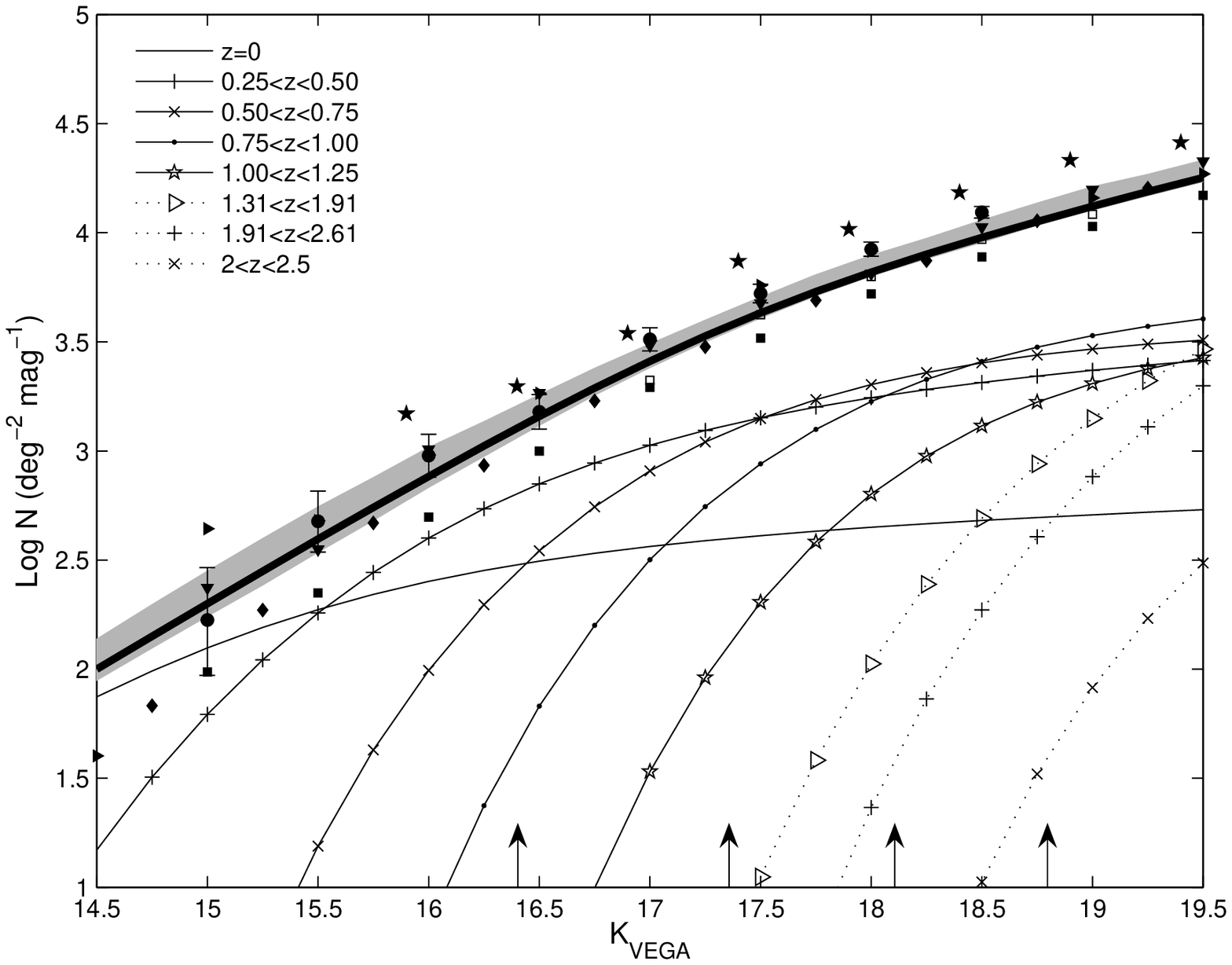}}
\resizebox{!}{0.40\hsize}{\includegraphics{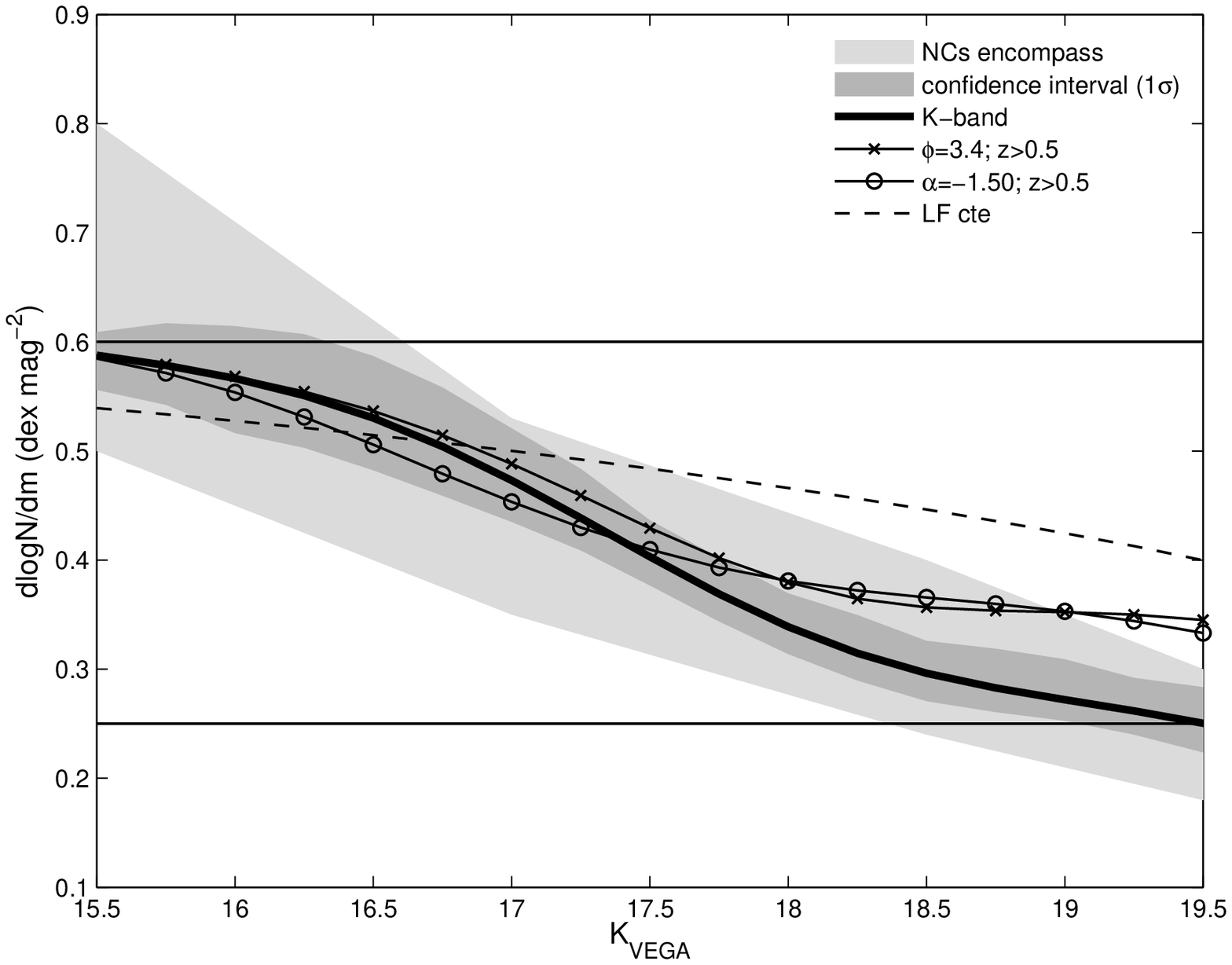}}
\caption{\label{slope_from_evolLF}\textit{Left:} $K$-band NCs derived from the LFs (thick black line) compared to data from other authors (the legend as in the left panel of Fig.~\ref{counts_slope_fields}). The shaded region indicates the $1\sigma$ confidence interval derived from the simulations on the uncertainties of the Schechter parameters. The continuous lines with different symbols show the redshift binned number counts at $\bar{z}=0.32,0.62,0.88,1.12$ derived from the LFs in the present work. The dashed lines with different symbols show the redshift binned number counts at higher redshifts from \cite{2006A&A...448..101G} (in the $i$ and $z$ rest-frame bands) and \cite{2007ApJ...656...42M} (in the R rest-frame band). These NCs have a negligible effect on the on the total NCs at $K$$<$18. The arrows depict the best-fit value of $M^{\ast}_{K,obs}$ for the LFs at $\bar{z}=0.32,0.62,0.88,1.12$. Note that to derive the total NCs we have also used LFs at higher redshift from the literature. \textit{Right:} Slope of the NCs derived from the LFs (thick black line). The dark grey shaded region represents the $1\sigma$ confidence interval derived from the simulations on the uncertainties of the Schechter parameters. The light grey shaded region shows the region encompassing the slopes derived from the references. The dashed line shows the predicted slope derived from a single LF constant through all redshifts. The circled line shows the predicted slope fixing $\alpha=-1.50$ for the LFs at $z>0.5$, The crossed line shows the predicted slope fixing $\phi^{\ast}_{K,obs}=3.4\times10^{-3}h_{70}^{3}Mpc^{-3}$ for the LFs at $z=0.5$.}
\end{figure*}

In the right panel of Fig.~\ref{evolving_param} we can see that the characteristic density, $\phi^{\ast}$, follows a decreasing trend in every band. The average estimates for optical LFs indicate a decrease of $\sim50\%$ in the number density from the local value to $z\sim1$ (\citealt{2006A&A...448..101G}, \citealt{2005A&A...439..863I}), whereas the evolution in the NIR LFs seems to be slightly weaker
($\sim30\%$). Also, the results in the $K$-band from \cite{2007arXiv0705.2438A} and \cite{2007MNRAS.380..585C} suggest a mild decreasing trend from $z=0.4-1.25$. Finally, the LFs of \cite{2007ApJ...656...42M} in optical bands confirm the
decreasing trend at high redshift finding a decrement of the
$\sim70-80\%$ in $\phi^{\ast}$ from the local value to z$\sim3$.

\subsection{LF evolution from the observed $K$-band}\label{LF_from_Kband}

The two panels of Fig.~\ref{evolving_param_Kobs} shows the multi-wavelength evolution of the Schechter parameters as probed by the observed $K$-band (thick black line); i.e, LFs at progressively shorter rest-frame wavelengths for growing redshifts. This is essentially the same plot as in Fig.~\ref{evolving_param} except that now the x-axis indicates the redshift where the $K$-band probes the different rest-frame bands. The vertical grey line centered at $z=1.25$ indicates the approximate redshift limit where the contribution to the $K$-band NCs ($K<19$) becomes smaller than $10\%$. Althought this means that the impact of the optical LFs on the bright counts is negligible, they illustrate the process of construction of NCs from LFs, allowing us to estimate the NCs to fainter magnitudes.

Note that in the optical bands, we have no references for the LFs at the precise redshift. Therefore, together with the estimated value of the $z$-band LF at $\bar{z}=1.42$ (open circle), derived from the Gabash et al. parametrization (see previous section), we show the measured value at $\bar{z}=1.7$ (black star), which presents an $M^{\ast}$ significantly fainter than their estimates at lower redshift. The same applies for the LF in the $i$-band at $\bar{z}=2.26$ (instead of $\bar{z}=1.89$) and the R-band at $\bar{z}=2.25$ (instead of $\bar{z}=2.36$).

\subsection{Summary}

In this section we presented the tools and measurements required to recover the NCs in terms of LFs at different bands and redshifts. As outlined at the beginning of this section, our motivation here is to determine the family of LFs at different redshifts that constitutes the NCs in the $K$-band, a result which is easily summarized in the thick black lines of Fig.~\ref{evolving_param_Kobs}.

As it can be seen from the left panel of that figure, the characteristic luminosity $M^{\ast}_{K,obs}$ shows an almost flat evolution beyond $z=0.4$. In fact, it is 1$\sigma$ compatible with a constant evolution:
$M^{\ast}_{K,obs}=-22.89\pm0.25$. On the other hand,
$\phi^{\ast}_{K,obs}$ shows a progressive decline of the
$\sim60\%$ from the local value in the rest-frame $K$-band to the
high redshift value in the R-band, with a significant decrease around $z\gtrsim1$.  A number of recent results seem to confirm the existence of this decline, suggesting that it is mostly driven by a steeper decrement in the number density of quiescent galaxies (\citealt{2006A&A...455..879Z}; \citealt{2007arXiv0705.2438A}; \citealt{2007ApJ...665..265F}). This trend is usually interpreted as an indicator of the beginning of an epoch of major build-up for these populations. Nevertheless, a deeper analysis of the evolution with redshift of the color bimodality is beyond the scope of this paper.

The evolution described by the $(K,obs)$ quantities has no direct physical meaning regarding galaxy evolution. However, it is crucial to describing how galaxy evolution is encrypted in the $K$-band NCs, and provides an interpretation for the shape of the NCs in terms of meaningful quantities, as we will show in the next section.

\section{$K$-band NCs from evolving LFs}\label{NCs from LFs}

Using the multi-wavelength evolution of $M^{\ast}_{K,obs}$, $\phi^{\ast}_{K,obs}$ presented in the previous section (thick black line of Fig.~\ref{evolving_param_Kobs}) and their corresponding values for the $\alpha$ parameter (see Table \ref{LF_summary}), we have derived the NCs in the $K$-band and its slope up to $K<19$ (these results are summarized in Table~\ref{Table_NCs_from_LF}). Additionally, we have calculated the $68\%$ confidence interval for both using Monte Carlo simulations on the Schechter function parameters. The random values of $M^{\ast}$ and $\alpha$ for each LF were taken from the error ellipses calculated with the STY method. When the ellipse parameters were not available, such as for the optical bands, we used a Gaussian distribution with a median equal to the given value and $\sigma$ equal to the 1$\sigma$ deviation.

In the left panel of Fig.~\ref{slope_from_evolLF} we compare the $K$-band NCs derived from the LFs (solid black line) with the observational results shown in the left panel of Fig.~\ref{counts_slope_fields} (the legend is the same). The remarkably good agreement proves the validity of the method applied to relate both quantities, and allows us to interpret the variation of the slope in terms of LFs. Therefore, we can consistently argue that the main causes of the shape of the $K$-band NCs are: the almost flat evolution with redshif of $M^{\ast}_{K,obs}$ and the significant decline of $\phi^{\ast}_{K,obs}$ with redshift. The former causes the redshift binned NCs to become progressively concentrated at fainter magnitudes due to the variation with redshift of the distance modulus (higher at low redshift but decreasing rapidly at higher redshifts). 
However, the dominant effect is the decrease of the characteristic density, which opposes the growing comoving volume. The relative growth in the comoving volume per redshift bin ($\Delta z=0.25$) progressively decreases from a factor of 2 at z$=$0.32 to 1.3 at z$=$1.12 (as we approach the peak at z$\sim$2). Simultaneously, $\phi^{\ast}_{K,obs}$ presents a decreasing trend that matchs the growth of the volume element at z$=$1.12, {\it freezing} the contribution of the LFs at higher redshifts, that otherwise would continue to increase to the peak of the volume element.

In terms of luminosity, it can be seen in the left panel of Fig.~\ref{slope_from_evolLF} that the decline of $\phi^{\ast}_{K,obs}$ leads to a smooth transition between the intrinsically faint galaxies at $z<0.5$ and the $\sim M^{\ast}_{K,obs}$ galaxies at $0.5<z<1$ around $K\sim17.5$ (the second and third arrows of the panel represent $M^{\ast}_{K,obs}$ at those redshifts). Hence, since the slope of the LFs (and therefore of the NCs) decreases rapidly after the knee, the transition leads to more flattening compared to the slope at $K=16$, where the $\sim M^{\ast}$ galaxies at $z<0.5$ clearly dominate the NCs.

To help us clarify the relevance of $M^{\ast}_{K,obs}$ and $\phi^{\ast}_{K,obs}$ for the modeling of the slope, in the right panel Fig.~\ref{slope_from_evolLF} we show the slope of the $K$-band NCs resulting from assuming the local $K$-band LF through all redshift bins (constant LF; dashed line). Despite the poor resemblance to the slope derived from the observed LFs (i.e., the slope of NCs in the left panel; solid black line), the decreasing trend is indicative that a gradual change in the slope is a natural consequence of the shape of the LFs modeled by the volume element. However, a precise evolution in the LFs is required to reproduce the change in the slope in the appropriate place.

To strengthen this idea we have also explored the scenario of no density evolution, fixing the value of $\phi^{\ast}_{K,obs}$ at $z>0.5$ (crossed line). In the absence of a decreasing characteristic density, the weigth of the M$\geq M^{\ast}$ galaxies up to z$\sim$2 would lead to a slope systematically larger than our 1$\sigma$ prediction at $K>17.5$, and clearly out of the envelope of the observed NCs at $K>18.5$.

Finally, we have tested the role of $\alpha$ in the shape of the NCs. As can be seen from both panels of Fig.~\ref{slope_from_evolLF}, the rather flat values of $\alpha$ in the low redshift LFs accentuates the decreasing trend of the slope. However, these values might be slightly underestimated due to the relatively shallow depth of the samples. The line with open circles in the right panel of Fig.~\ref{slope_from_evolLF} shows the predicted slope with $\alpha=-1.50$ fixed in the LFs at $z>0.5$. As expected, the steeper faint end does not significantly affect the NCs around $K\sim17.5$, which are dominated by $M^{\ast}$ galaxies. However, it does predict a higher slope at fainter magnitudes that falls outside the region defined by the observations. This test poses a restriction against high values of $\alpha>-1.5$ in the low redshift (z$=$0.5-1) NIR LFs.

\begin{table}[t]
\centering
\small
\begin{tabular}{cccc}
\hline
K Bin Center & log(N) & dlog(N)/dm \\
\hline
      15.0000&    2.3005&    0.6013\\ 
      15.5000&    2.5959&    0.5877\\
      16.0000&    2.8835&    0.5666\\
      16.5000&    3.1584&    0.5301\\
      17.0000&    3.4114&    0.4730\\
      17.5000&    3.6327&    0.4030\\
      18.0000&    3.8193&    0.3388\\
      18.5000&    3.9783&    0.2962\\
      19.0000&    4.1209&    0.2720\\
      19.5000&    4.2530&    0.2504\\
\hline
\end{tabular}
\caption{\label{Table_NCs_from_LF} Theoretical $K$-band NCs and NCs slope in 0.5 magnitude bins derived from the LF of section \ref{RESULTS}.}
\end{table}

\subsection{Summary and predictions for fainter magnitudes}
Within the three regime schema for the evolution of the slope proposed in section $\S$\ref{MODEL}, it can be seen that the slope of the $K$-band NCs leaves the Euclidean limit around $K\sim15.5$. It reachs the ($M^{\ast}$ dominated) transition regime faster (K$\sim$17.5) than in a no evolution scenario (K$\geq$19) due to a significant decrease in $\phi^{\ast}_{K,obs}$ that {\it freezes} the contribution from bright $z>$1 galaxies. The migration into the $\alpha$ regime is harder to determine. The relatively low ($\alpha=-0.9,-1$) faint end slope at z$\sim$1 might cause LFs with larger values of $\alpha$ to dominate the faint counts. These LFs would not be severely suppressed if the decreasing trend in $\phi^{\ast}_{K,obs}$ stabilizes at $z>1.5$, given that the volume element evolves smoothly between z$=$1.5-4. Therefore, for any $\alpha<$-1 the high-z LFs could eventually dominate the NCs at sufficiently faint magnitudes.

For the set of LFs that we have compiled, Gabash et al. propose a fixed $\alpha=$-1.33 for the LFs at z$=$1.3-2.6 in the $i,z$ bands, whereas Marchesini et al. give $\alpha=-1.07$ in the R-band at z$=$2.35. Interestingly, several authors report significantly larger values of $\alpha$ for the optical LFs at high-z (\citealt{2007ApJ...656...42M},  $\alpha=$-1.40 in the B-band at z=3; \citealt{2008ApJS..175...48R}, $\alpha=$-1.80 at 1700$\AA$ at z=3). 
Based on these numbers, we predict a slowly decreasing slope from $d\log N/dm\sim0.3-0.2$ in the magnitude range $K=$19.5-22, due to the dominant contribution from the z$\sim$2 bin ($\alpha$=-1.33). However, the slope in this range is not purely in the asymptotic $\alpha$ regime (that would lead to $d\log N/dm\sim0.13$). The resulting value is a weighted sum of slopes which includes a contribution from $z>2$ (around $K\sim20$) $M^{\ast}$ galaxies. This contribution from larger slopes lowers the rate of approach to the asymptotic limit. Furthermore, if the slope of the high-z LFs grows rapidly to large values ($\alpha\sim-1.8$) we could measure an increase in the slope around $K>$23. 

\subsection{NCs in other NIR bands}

The procedure to derive NCs from LFs described in section \ref{RESULTS} can be used to reproduce the NCs in any other band using the appropriate LFs and redshifts. Nevertheless, based on the multiwavelength LFs probed by the observed $K$-band (summarized in Fig.\ref{evolving_param_Kobs} for the $K$-band) it is possible to predict the approximate shape of the NCs in the closest NIR bands; e.g., IRAC-3.6 or H (1.65$\mu m$). 

Firstly, if the NCs in different bands are referred to AB magnitudes, the results can be compared in a more meaningful context, avoiding the offset introduced by the AB-Vega transformation. In the AB-system, the magnitude range where the NCs fall in the Euclidean regime ($d\log N/dm\sim$0.6) is determined by the local value of $M^{\ast}$ and $\phi^{\ast}$ in that band. Therefore, similar values of the local LFs will result in similar NCs at bright magnitudes. 

Furthermore, if the multiwavelength evolution of $M^{\ast}$ and $\phi^{\ast}$ probed by the different bands exhibits a similar evolution, the slope of the NCs should change around the same magnitude range. That is the case of the $H$ and $K$ bands. At z$=$0, we have derived the H-band LF directly from the $K$-band applying a color term. At higher redshifts, the multiwavelength LFs probed by the observed $H$-band ($J$-band at z=0.32 and z-band at z=0.83) presents similar values to the LFs probed by the $K$-band at the same redshifts (see Fig.~\ref{evolving_param_Kobs}). As a consequence, the NCs in the $H$-band also present a flattening in the slope around m[AB]$\sim$19-19.5 (see \citealt{2006MNRAS.370.1257M} and references therein). In addition, \cite{2008MNRAS.386...11M} have shown that the NCs in the IRAC-3.6 band exhibit a significant flattening at slightly fainter magnitudes (m[3.6]$\sim$20). This suggests that IRAC-3.6 and K probe a similar evolution in the LFs. If we estimate the [3.6] local LF following the same approximation as for the $H$-band, the shift in the slope flattening is consistent with applying the typical color K-[3.6]$\sim$-0.6~mag (\citealt{2007ApJ...655..863D}) to the $K$-band local LF. 

A detailed study of the change in the slope would depend on the multi-band LFs probed by the given band. Nevertheless, this simple calculation provides a rough estimate of the magnitude where the change in the slope begins to be appreciable.

\section{Summary and conclusions}\label{CONCLUSIONS}

In this paper we present the data of a NIR photometric survey in
the HDFN and Groth fields. Combining this data with the
panchromatic sets available we have extracted a $K$-band selected sample up to $K=18.5$ in a combined area of $\sim0.27$~deg$^{2}$. Additionally, we made use of the deep NIR public survey conducted by ESO in the CDFS ($172$~arcmin$^{2}$) to complement our relatively shallow survey and increase the surveyed area. We have derived high quality photometric redshifts for the whole sample, taking advantage of the excellent coverage of the SED. $K$-band galaxy NCs have been derived in the three fields covering the range $16<K<18.5$. The comparison to a compilation of shallow wide-area and deep pencil-beam surveys shows a general good agreement in spite of the considerable scatter, a consequence of the cosmic variance. We have studied the redshift distribution of the NCs finding that the $90\%$ of the galaxy counts up to $K\sim18.5$ come from the low-redshift population ($z<1$). Additionally, we have measured the effects of cosmic variance by comparing the redshift binned NCs in our three fields, finding that field-to-field differences reach $40\%$ at a certain magnitude, with consequent impact on the NC distribution. 

We have provided evidence of an average decreasing trend in the slope of the $K$-band NCs, $\sim50\%$ in the $15.5<K<18.5$ range, evolving from $d\log N/dm=0.6-0.3$. The comparison to  100 synthetic catalogs from the Millennium simulation reveals that cosmic variance in areas $\sim0.25$deg$^{2}$ leads to significant scatter in the observed flattening rate of the slope. 

We have studied the composition of the NCs in terms of LF building blocks, concluding that the change in the slope with observed magnitude can be summarized in three main regimes. Firstly, at bright magnitudes, the classical Euclidean regime ($d\log N/dm=0.6$) is dominated by low redshift $M^{\ast}$ galaxies. Then, at intermediate magnitudes, the transition regime is dominated by the LF at the redshift that maximizes $\phi^{\ast}\frac{dV_{c}}{d\Omega}$. Here the slope decreases rapidly around the apparent magnitude of $M^{\ast}$ at that redshift. Finally, at faint magnitudes, the ``$\alpha$ regime'' is populated by galaxies at the faint end of a combination of LFs. Here the slope asymptotically approaches a minimum value at $\sim$-0.4(1+$\alpha$). The value of $\alpha$ will typically be given by the LF at the maximum of $\phi^{\ast}\frac{dV_{c}}{d\Omega}$ or a close LF with a much larger faint end slope.

We have explored the evolution of the $K$-band NCs by deriving LFs in the observed $K$-band as a function of redshift (at a mean rest-frame wavelength $2.16\mu m/(1+\bar{z})$), and complementing our data with optical and NIR LFs from the literature. In terms of the multi-wavelength LFs, we find that the flattening of the slope is the consequence of a prominent decrease of the characteristic density $\phi^{\ast}_{K,obs}$ ($\sim60\%$ from $z=0.5$ to $z=1.5$) and the almost flat evolution of $M^{\ast}_{K,obs}$ (1$\sigma$ compatible with $M^{\ast}_{K,obs}=-22.89\pm0.25$). The combination of both effects forces a transition to the $\alpha$ regime at $K\sim17.5$ that otherwise would have taken place $1-2$ magnitudes later. 

Our predictions for $K\gtrsim20$, based on LFs from the literature, suggest that the slope will continue to decrease steadily to $d\log N/dm\lesssim$0.2 dominated by intrinsically faint galaxies from a mixture of LFs at z$=$1.5-2 and a minor contribution from bright galaxies at z$>$2. However, if optical LFs at $z>3$ present an even larger faint end, as suggested by some authors, there might be a slope increase to a higher asymptotic limit around $K\gtrsim22.5$.

\section*{Acknowledgments}
We thank the referee M. Bershady for his useful and constructive comments.
We acknowledge support from the Spanish Programa Nacional de
Astronom\'{\i}a y Astrof\'{\i}sica under grant AYA 2006--02358. Partially funded by the Spanish MEC under the Consolider-Ingenio 2010 Program grant CSD2006-00070: First Science with the GTC (http://www.iac.es/consolider-ingenio-gtc/)
Based on observations collected at the Centro Astron\'{o}mico Hispano Alem\'{a}n (CAHA) at Calar Alto, operated jointlyby the Max-Planck Institut fur Astronomie and the Instituto de Astrof\'{i}sica de Andaluc\'{i}a (CSIC). 
This article is based on observations made with the WHT operated on the island of La Palma by the Instituto Astrof\'{\i}sico de Canarias in the Spanish Observatorio del Roque de los Muchachos. This work is based in part on observations made with the {\it Spitzer} Space Telescope, which is operated by the Jet Propulsion Laboratory, Caltech under NASA contract 1407. GALEX is a NASA Small Explorer launched in 2003 April. We gratefully acknowledge NASA's support for construction, operation, and scientific analysis of
the GALEX mission. Based in part on data collected at Subaru Telescope and obtained from the SMOKA, which is operated by the Astronomy Data Center, National Astronomical Observatory of Japan. This publication makes use of data products from the Two Micron All Sky Survey, which is a joint project of the University of Massachusetts and the Infrared Processing and Analysis Center/California Institute of Technology,
funded by the National Aeronautics and Space Administration and the National Science Foundation. Based on observations obtained with MegaPrime/MegaCam, a joint project of CFHT and CEA/DAPNIA, at the Canada-France-Hawaii Telescope (CFHT) which is operated by the National Research Council (NRC) of Canada, the Institut National des Science de l'Univers of the Centre National de la Recherche Scientifique (CNRS) of France, and the University of Hawaii.

\bibliography{referencias}
\bibliographystyle{aa}
\end{document}